\documentclass{svjour3}                     
\smartqed  
\usepackage{graphicx}
\begin{document}

\title{Fresnel Zone Plate Telescopes for X-ray Imaging I: Experiments with 
a quasi-parallel beam \thanks{This experiment was made possible in part by a grant from
Indian Space Research Organization to ICSP. SP and DD thank CSIR/NET scholarships which supported
their research work.
}
}

\titlerunning{Zone Plates for X-ray Imaging}   

\author{Sandip K. Chakrabarti \and S. Palit \and D. Debnath \and A. Nandi \and V. Yadav 
}

\authorrunning{Chakrabarti et al. } 

\institute{Sandip K. Chakrabarti \at
              S.N. Bose National Centre for Basic Sciences, JD Block, Salt Lake, Kolkata 700097 \\
(Also at Indian Centre for Space Physics, 43 Chalantika, Garia Station Rd., Kolkata 700084)\\
              Tel.: +91-33-23355706\\
              Fax: +91-33-23353477\\
              \email{chakraba@bose.res.in}           
           \and
            S. Palit, D. Debnath, A. Nandi$^+$, V. Yadav$^+$ \at
Indian Centre for Space Physics, 43 Chalantika, Garia Station Rd., Kolkata 700084\\
Tel.: +91-33-24366003\\
              Fax: +91-33-24622153 Ext. 28\\
              \email{sourav@csp.res.in; dipak@csp.res.in; anuj@csp.res.in; vipin@csp.res.in}   
($+$: Posted at ICSP by Space Science Division, ISRO Head Quarters)
}

\date{Received: date / Accepted: date}

\maketitle

\begin{abstract}
Combination of Fresnel Zone Plates (FZP) can make an excellent telescope for
imaging in X-rays. We present here the results of our experiments with several
pairs of tungsten made Fresnel Zone plates in presence of 
an X-ray source kept at a distance of about $45$ feet. The quasi-parallel beam
allowed us to study sources placed on the axis as well as off the axis of the telescope.
We present theoretical study of the fringe patterns produced by the zone plates
in presence of a quasi-parallel source. We compare the patterns obtained 
from experiments with those obtained by our Monte-Carlo simulations.
The images are also reconstructed by deconvolution from both the patterns. We compare the 
performance of such a telescope with other X-ray imaging devices used in space-astronomy.

\end{abstract}

\section{Introduction}
\label{intro}

Imaging in X-rays in space has always been a challenging task. X-ray films are generally very inefficient and
in space it is inconvenient to take continuous imaging with X-ray films. To circumvent this, Coded
Aperture Masks (CAM) (Mertz \& Young, 1960; Ables, 1968; Dicke, 1968) are widely used. 
While a CAM has an advantage that it is a single element instrument,
it has the disadvantage that the resolution is low, and is limited by the smallest mask element.
Recently,  grazing incidence focusing instruments are being developed 
specially in space missions such as NUSTAR and SIMBOL-X (Ramsay, 2006; Pereschi, 2008), 
where target resolutions of $20-40$ arc seconds
are being contemplated. A relatively simpler concept is, however, present in the 
literature. It was realized (Mertz 1965) that high resolution can be achieved 
when X-ray shadows are cast with two perfectly aligned Fresnel zone plates and 
the Moir\'e fringes could be discerned by suitable detectors. Furthermore, the resolution is
independent of the energy of X-rays. Theoretically, the resolution could be arbitrarily high
as it increases with the separation $D$ of the zone plates, apart from being proportional to the
finest element in the plate. Since small-pixel detectors were unavailable even though this
concept was well established, there was hardly any application. Desai and his collaborators
(1993, 1998, 2000) showed renewed interests in this type of telescopes and
actually carried out an experiment and showed that an image reconstruction was possible. 

However, so far, a detailed and quantitative exploration of imaging with single and multiple sources 
in laboratory environments has not been carried out with Fresnel Zone Plate telescopes.
With the recent development of CMOS based X-ray detectors where each pixel could be 
as small as $50$ microns, it has become necessary to revisit the problem once more to actually study the 
feasibility of sending Zone Plate telescopes for space applications. Furthermore, the high resolution
imaging devices could be used for medical purposes also. This is important since a high dose 
of X-ray is required when conventional X-ray films are used. In laboratory and medical uses, 
however, the beams are likely to be diverging. It is therefore necessary to study the 
feasibility of imaging with diverging beams as well.

In this series of papers, we plan to explore the practical aspects of the zone plate telescopes,
focusing on the effects of the divergence of the beam and achievable resolutions in diverging beams.
We also study with various combinations of zone plates. Our results are also supported by Monte-Carlo
simulations carried out with parallel and divergent beams. Later in this series we
will describe the results of the test and evaluation of two zone plate telescopes for the RT-2
experiments on board the Russian satellite CORONAS-PHOTON to be launched shortly.

The plan of our present paper is the following. In the next Section, we briefly describe the mathematical 
formalism behind the functioning of a Fresnel zone plate telescope and the properties of the 
Moir\'e patterns falling on the detector. In Sections
3 and 4, we describe the setup of our experiment and the scheme of Monte-Carlo simulations. 
In Section 5, we present results obtained with a single on-axis or off-axis source
with two different configurations of the zone plates and with two types of zone plates.
Monte-Carlo simulations of these cases are also presented to support the experimental 
results. In Section 6, we present results with multiple sources and present simulation results.
Finally, in Section 7, we make concluding remarks. In Paper II, of this series we will 
make a comparative study of Monte-Carlo simulations for single and multiple sources at various 
distances as well as for parallel beam, which is not achievable in the laboratories. 
In Paper III (Nandi et al., 2008), we will discuss the 
test and evaluations of the RT-2 payload to be launched aboard CORONAS-PHOTON satellite.

\section{Mathematical Formalism of the Zone Plate Optics}

\subsection {Source located at a large distance}

If $f({\vec r})$ is the source function, then the 2-dimensional Fourier transform
is defined by,
$$
F({\vec \rho}) = \int_\infty f({\vec r}) exp(-2\pi i {\vec \rho} . {\vec r}) d^2 {\vec r},
\eqno{(1)}
$$
where $\vec \rho$ is the spatial frequency for which the transformation is 
computed. This can be re-written as,
$$
F(\vec \rho) = 
\int_\infty f({\vec r}) \frac{1}{2} [1+cos(2\pi{\vec \rho}.{\vec r})] d^2 r 
-\int_\infty f({\vec r}) \frac{1}{2} [1-cos(2\pi{\vec \rho}.{\vec r})] d^2 r 
$$
$$
\ \ \ \ \ \ \ \ \ +i \int_\infty f({\vec r}) \frac{1}{2} [1+sin(2\pi{\vec \rho}.{\vec r})] d^2 r 
-i \int_\infty f({\vec r}) \frac{1}{2} [1-sin(2\pi{\vec \rho}.{\vec r})] d^2 r  .
\eqno{(2)}
$$
Here, the exponential term of eq. (1) has been broken in a manner that each transmittance
term on eq. (2) can have non-negative values only. Each such term represents a mask. Two sets have 
positive transmittance and two sets have negative transmittance. Two of them will produce cosine 
transformation and other two will produce sine transformation (Barrett \& Myers, 2004).

In order to produce a suitable mask having above transmittance, we recall that a 
Fresnel zone plate has the following transmittance function,
$$
T(r) =  \frac{1}{2}\{1 \pm sgn [sin(\alpha r^2)]\} S({\vec r}) ,
\eqno{(3a)}
$$
where $\alpha$ is a constant and $S({\vec r})$ is a support function (equals to $1$ within 
the outer boundary of the zone plate and zero outside). The sgn function is $+1$ 
when sin$(\alpha r^2) > 0$ and $-1$ when  sin$(\alpha r^2) < 0$. The transition from opaque to 
clear zones occurs at the zeros of sin($\alpha r^2$), i.e., at $\alpha r_n^2= n\pi$.
Here, $r_1 = \pi/\alpha$ is the radius of the innermost zone and $r_n=\sqrt{n} r_1$. It is also 
possible to have a zone plate having transmittance function 
$$
T_{ZP}({\vec r})= \frac{1}{2}\{1 \pm sgn [cos(\alpha r^2)]\} S({\vec r}) .
\eqno{(3b)}
$$
Here, the sign of the cosine term changes at $\alpha r_n^2=(n\pm\frac{1}{2})\pi) $
for $n=1,2,3 ... ...$.

In order to see how a set of zone plates, each having a transmittance given by
either eq. (3a) or eq. (3b) (with either positive or negative sign) may behave as a Fourier transformer
as presented in eq. (2), we expand the left hand side of eq. (3a) as,
$$
T_{ZP}({\vec r}) = [\frac{1}{2} + \frac{1}{i\pi} \Sigma_{k=- \infty,\ k=odd}^{k=\infty}
\frac{1}{k} exp ( - i \alpha k r^2 ) ] S({\vec r}) .
\eqno{(4)}
$$

Combining with another zone plate of transmittance $T({\vec r} - {\vec r_s})$ and expanding in a similar 
series with $k'$ replacing $k$, and keeping only the lowest terms ($k=-k'=\pm 1$), we obtain, 
$$
T_{ZP1}({\vec r}) T_{ZP2}({\vec r} - {\vec r_s}) = [\frac{1}{4} + \frac{2}{\pi^2} cos(2\alpha {\vec r}.{\vec r_s}
-\alpha r_s^2)] S({\vec r}, {\vec r_s}) + {\rm  other \  terms}
\eqno{(5)}
$$
where $S({\vec {r}}, {\vec {r_s}}) = S({\vec {r}})S({\vec {r_s}})$.
Here, ${\vec r_s}$ being the shift vector in the plane of the zone plate of the origin of the second 
plate with respect to the first. By choosing the lowest $k$ we ignored the lesser significant higher
order terms, chirp terms, odd harmonics terms etc. (Shulman, 1970; Barrett \& Swindell, 1981, 1996).
The `other terms' include these and also the original transmittance of each zone plate. 

It is clear that the bright Moir\'e fringes will be produced when the argument of cosine will be unity, i.e.,
$\alpha {\vec r}.{\vec r_s} -\alpha r_s^2 = n\pi$ with $n=1,2,3 ... ...$. First note that
since ${\vec r}_s$ and $\alpha$ are constants, the projection of ${\vec r}$ on ${\vec r}_s$ must be fixed
on a given bright fringe, i.e., on a constant $n$. Thus the fringes must be straight
lines. Second, the fringe separation is $\pi/{\alpha r_s}$. Thus, when the relative shift
${\vec r}_s$ becomes larger, the fringe separation becomes smaller. 

In a zone plate telescope, plates are aligned but the source is kept off-axis in order to 
get relative shift ${\vec r}_s$ of the ray of light. Sources at a higher off-axis location produce
finer fringe separations. Whether these are detectable depends on how fine the detector 
pixels are. The fringe patterns on the detector are inverse Fourier transformed 
to get the source location. 

It is easy to analysis the results for different combinations of the zone plates. We summarize 
a few interesting ones here.

\noindent Case (i): When both the plates have positive transmittance with zone boundaries at $r_n = \sqrt{n} r_0$ or
$r_n=\sqrt{(n+\frac{1}{2})}r_0$ ($r_0$ is a constant.), the combined transmittance is: 
$$
T_{ZP1}({\vec r}) T_{ZP2}({\vec r}-{\vec r_s})=[\frac{1}{4}+\frac{2}{\pi^2}cos(2\alpha {\vec r}.{\vec r_s}-
\alpha r_s^2) ] S({\vec {r}}, {\vec {r_s}}).
\eqno{(6a)} 
$$
This is a cosine transformer and the brightness modulation of the fringe is between $\frac{1}{4} +\frac{2}{\pi^2}$
and $\frac{1}{4} - \frac{2}{\pi^2}$.

\noindent Case (ii): When two plates have positive transmittance and one has zones at $r_n = \sqrt{n} r_0$ 
while the other has zones at $r_n=\sqrt{(n+\frac{1}{2})}r_0$, the combined transmittance is: 
$$
T_{ZP1}({\vec r}) T_{ZP2}({\vec r}-{\vec r_s})=[\frac{1}{4}+\frac{2}{\pi^2}sin(2\alpha {\vec r}.{\vec r_s}-
\alpha r_s^2) ] S({\vec {r}}, {\vec {r_s}}).
\eqno{(6b)} 
$$
This is a sine transformer and the brightness modulation is same as above.

\noindent Case (iii): When one plate has a positive transmittance and the other has a negative
transmittance and both have zones at $r_n = \sqrt{n} r_0$ or at $r_n=\sqrt{(n+\frac{1}{2})}r_0$, 
the combined transmittance is: 
$$
T_{ZP1}({\vec r}) T_{ZP2}({\vec r}-{\vec r_s})=[\frac{1}{4}-\frac{2}{\pi^2}cos(2\alpha {\vec r}.{\vec r_s}-
\alpha r_s^2) ] S({\vec {r}}, {\vec {r_s}}).
\eqno{(6c)} 
$$
This is a cosine transformer and the brightness modulation is the same as above, but the fringes 
are shifted on the detector plane by a phase shift of $\pi$.  

\noindent Case (iv): When one plate has a positive transmittance and the other has a negative
transmittance and one has zones at $r_n = \sqrt{n} r_0$ and the other has zones at 
$r_n=\sqrt{(n+\frac{1}{2})}r_0$, the combined transmittance becomes:
$$
T_{ZP1}({\vec r}) T_{ZP2}({\vec r}-{\vec r_s})=[\frac{1}{4}-\frac{2}{\pi^2}sin(2\alpha {\vec r}.{\vec r_s}-
\alpha r_s^2) ] S({\vec {r}}, {\vec {r_s}}).
\eqno{(6d)}
$$
This is a sine transformer and the brightness modulation is the same as above, but the fringes
are shifted with respect to eq. (6b) on the detector plane by a phase shift of $\pi$.

If the experiment is carried out with either cosine or sine transformers, and the resulting 
shadows are inverse Fourier transformed following procedures presented in, say, Mertz (1996), 
there would be symmetrically placed pseudo images and 
also a dc signal at the center (due to the $1/4$ term in each transmittance). 
If the combinations (i) \& (ii) or (iii) \& (iv) are used, the dc signals will remain, though the 
pseudo images will disappear. Only when all four sets are used, would the source be reproduced 
accurately. However, since many terms are ignored, there would be faint traces from higher harmonics.

The `ideal' resolution of the zone plate telescope is easy to determine: this is dictated by the 
highest possible angular separation of two sources, photons from which pass through the same transparent zone
while reaching the same point on the detector plane (assumed to be placed immediately behind 
the second zone plate). Only when they are separated beyond an angular separation 
of  $r_{N}-r_{N-1}/D$, where $N$ is the
total number of zones in a plate), the ray from one will pass through the transparent zone and the 
other through the opaque zone. Thus the ideal resolution is $r_{N}-r_{N-1}/D$ radian. The resolution becomes
twice as bad since there are two zone plates. This will be further limited by the detector
pixel size, in case it is larger that the thinnest zone. This resolution is possible only if the 
source is located at infinity. When the source is at a finite distance, the resolution will be 
further deteriorated due to fact that the projected shadows of the zones are larger.
We shall discuss these issues and demonstrate this in Paper II. However,
in passing we wish to point out that by increasing $D$ we can increase the resolution to be
as high as desired provided the pointing accuracy is also high enough.

\subsection{Source located at a finite distance}

Since for any laboratory experiment we must have the source at a finite distance, the beam
can not be quite parallel. It is important to understand how the result is affected by the 
divergence. When the source is located at a finite distance, the computational procedure is similar,
except that the fringes are no longer straight line in nature. This is shown below.
Fig. 1 shows the cartoon diagram of the setup. ZP1 and ZP2 are the zone plates separated by a distance of D along the Z-axis of the 
coordinate system. The source S is located at the radial vector ${\vec R}$ in the X-Y plane 
at  a distance of $z$ from the first zone plate. The ray passing through an
arbitrary point P$_1$ of ZP1 will intercept ZP2 at P$_2$ while its projection on
ZP2 will be at Q. The projection of the source S on ZP1 is at S$_1$. If S$_1$P$_1$ makes an 
angle $\phi({\vec r})$ at S, then QP$_2$ will also make $\phi({\vec r})$ at P$_1$. 

\begin{figure}[h]
\includegraphics[height=2.15in]{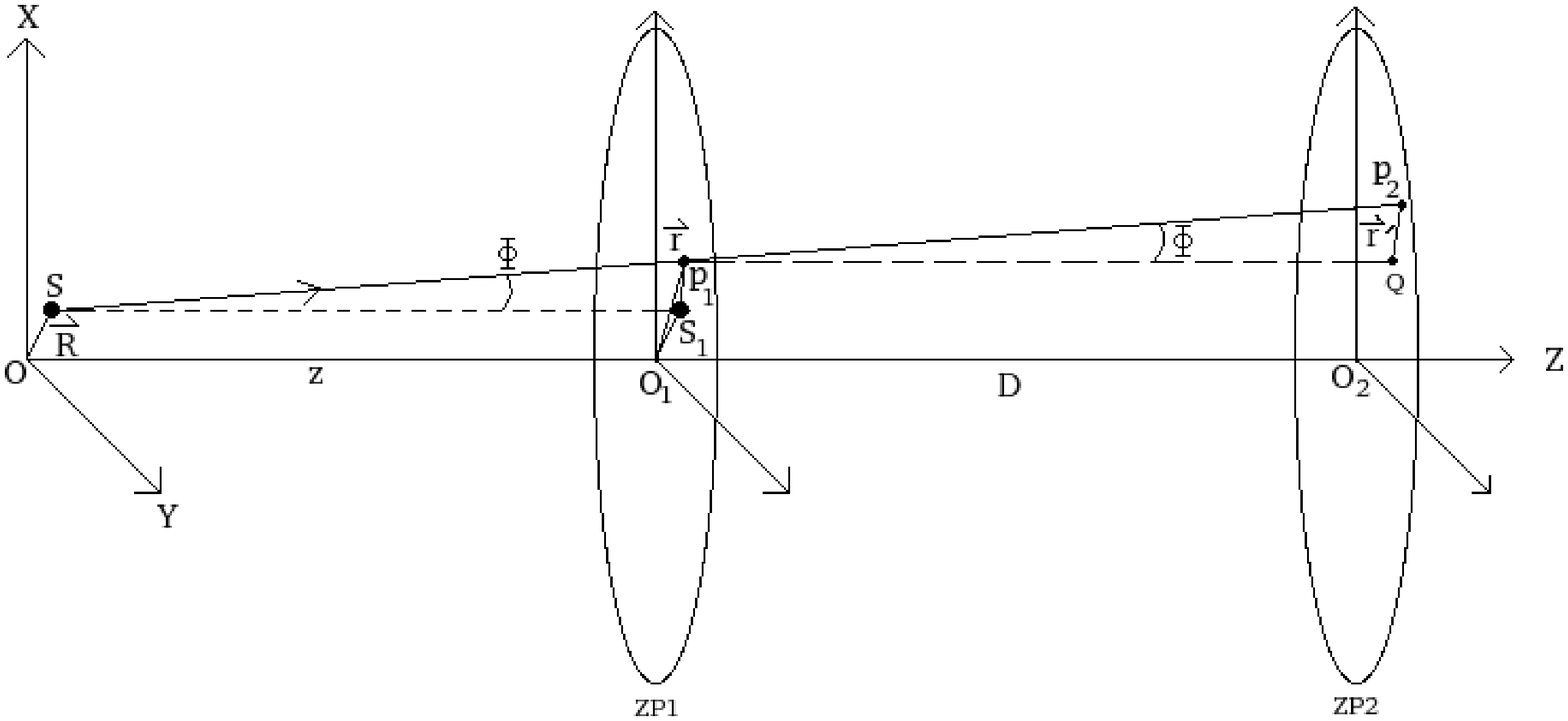}
\caption{A cartoon diagram of the setup to compute the combined shadow geometry. Here $z$ is the
distance of the point source S from the first zone plate (ZP1) and D is the 
distance between the two zone plates, ZP1 and ZP2. $P_1$ and $P_2$ are the
points were a ray intercepts the zone plates. We assumed the thickness
of the zone plates to be negligible for simplicity of computation of the
resulting shadow patterns.}
\label{}
\end{figure}

From Fig. 1, we must have, 
$$
\frac {S_1P_1}{z}=\frac{{\vec r} -{\vec R}}{z} = \frac {QP_2}{D} =\frac{\vec r'}{D} = {\rm tan} \Phi ({\vec r}) .
\eqno{(7)}
$$
If we follow the same analysis as in \S 2.1, we find that the combined transmittance in this case is given by,
$$
T_{ZP1}({\vec r}) T_{ZP2}({\vec r} - {\vec r'}) = [\frac{1}{4} + \frac{2}{\pi^2} cos(2\alpha {\vec r}.{\vec r'}
-\alpha r'^2)] S({\vec {r}}, {\vec {r'}}) + {\rm  other \ terms}
\eqno{(8)}
$$
Exactly in the same way as before, if we concentrate on the brightest fringes, we 
must have,
$$
2\alpha {\vec r}.{\vec r'} -\alpha r'^2 = {\rm const.} = 2n\pi \ \  (\rm n=1,2,3 ... ...) .
\eqno{(9a)}
$$
Substituting ${\vec r'}$ from eq. (7) and expanding we get,
$$
r^2 - {\vec r}.{\vec R} (1 - \frac{D}{2z}) = \frac{2n\pi}{A},
\eqno{(9b)}
$$
where $A=\frac{2 \alpha D}{z} (1-\frac{D}{2z})$. Here, we made an assumption that 
$1-\frac{D}{z} \sim (1- \frac{D}{2z})^2 $, since, normally $z>>D$. Putting 
$r = \hat i x + \hat j y$ and ${\vec R}= \hat i x_S + \hat j y_S$, and re-arranging, we
obtain,
$$
[x-\frac{x_S}{2}(1-\frac{D}{2z})]^2 + [y-\frac{y_S}{2}(1-\frac{D}{2z})]^2  
= \frac{2n\pi}{A} - \frac{1}{4} (1-\frac{D}{2z})^2 R^2.
\eqno{(10)}
$$
We observe that for a generally placed off-axis point source, located at a finite distance $z$ from ZP1, the 
brightest rings form concentric circles with radii dependent on $R$. In case of an on-axis source at a finite 
distance, we put $x_S=0=y_S$. The radii of the brightest circular fringes would then be located at 
$$ 
x^2 + y^2 = \frac{2n\pi}{A} .
\eqno{(11a)}
$$
and the darkest circular fringes would be located at,
$$ 
x^2 + y^2 = \frac{n\pi}{A} .
\eqno{(11b)}
$$
where $n$ is a positive integer. 
It is clear that the separation between the concentric circles will be reduced as $n$ becomes larger.

In passing, we may mention that when we have four sets of zone plates in a telescope, 
the vectors pointing to the source S from the 
origin of the whole telescope will not in general be located symmetrically about the four separate
axis of the four sets of zone plates. In other words, the vectors ${\vec R}_1$, ${\vec R}_2$, 
${\vec R}_3$ and ${\vec R}_4$  pointing to the source from the four optical axis 
would in general be unequal. Thus separate sets of concentric
circles will be produced for each set and their centers need not be located at the same point in the
detector plane.

When the source is far from the axis, higher $n$ concentric circular fringes will fall on the 
detector plane. Strictly speaking, the fringes will still not be straight lines. One can recover 
straight fringes only when $z \rightarrow \infty$.

\section{Experimental setup}

We have experimented with various types of zone plates.

\noindent Type A: This has $n_{th}$ radius located at $r_n=\sqrt{n} r_1$. The central
zone of radius $r_1$ is transparent  (positive transmittance)
and $r_1=0.122$ cm. There are altogether 150 zones and the smallest zone width is 
$50$ micron. Thickness of the whole plate is $1$ mm. When two such plates are kept separated at
a distance of $20$ cm, the angular resolution becomes $\sim 104$ arc seconds. 

\noindent Type B: This has $n_{th}$ radii at $r_n=\sqrt{n} r_1$. The central
zone of radius $r_1$ is transparent and $r_1=0.1$ cm. There are altogether 144 zones and the smallest zone is 
$41.74$ micron. Thickness of the whole plate is $1$ mm. When two such plates are kept separated at
a distance of $20$cm, the angular resolution becomes $\sim 86 $ arc seconds. 

\noindent Type C: This has $n_{th}$ radii at $r_n=\sqrt{n\pm{1/2}} r_0$, where $r_0$ is a constant. The central
zone of radius $r_1$ could be opaque or transparent and $r_0=0.1$ cm. There are 
altogether 144 zones and the smallest zone is $41.66$ micron. The thickness of the whole plate is $1$ mm. 
When two such plates are kept separated at a distance of $20$ cm, the angular resolution becomes $\sim 86$ 
arc seconds. 

In Figs. 2(a-b) we show images of zoomed in photos of such zone plates taken on X-ray
films in our laboratory. The smallest fringes of $\sim 40$ micron can be seen very clearly.
Unfortunately, for imaging reconstruction it is difficult to use photographs on films as the
digitization is required.  Furthermore, there would be noise introduced  during the 
developing process. For this, we use the Rad-Icon made RadEye CMOS detector camera 
having a minimum pixel size of $50$ micron. This is also the minimum fringe 
separation that we can resolve. This therefore restricts imaging sources
to be too far away from the optical axis. 

Fig. 3a shows our experimental setup. It consists of an X-ray generator with a molybdenum target. The beam
line is about 45 feet long. The anode voltage and the current can be controlled. 
At the end of beam-line the tungsten zone plates are placed, co-aligned in an aluminum tube. We made 
aluminum tubes,  of $10$ cm and  $20$ cm in length. Since the zone
plates we use are made up of $1$ mm thick tungsten, the `opaque' zones are opaque up to $\sim 100$keV.
Thus we did not choose to vacuum the beam-line and supply high energy X-rays by 
applying higher anode voltage instead. The intervening air column will cause absorption in the soft X-rays but 
hard X-rays including the K$_\alpha$ and K$_\beta$ lines will remain strong. 
In Fig. 3b we show the setup at the detector end of the X-ray beam. Here, an aluminum tube 
holds zone plates in place and the CMOS imager are shown. We used both $10$ cm and $20$ cm 
long tubes which separated the plates. The digital images are captured by software
on the computer screen (not shown) before further processing.

\begin{figure}[h]
\includegraphics[height=1.80in]{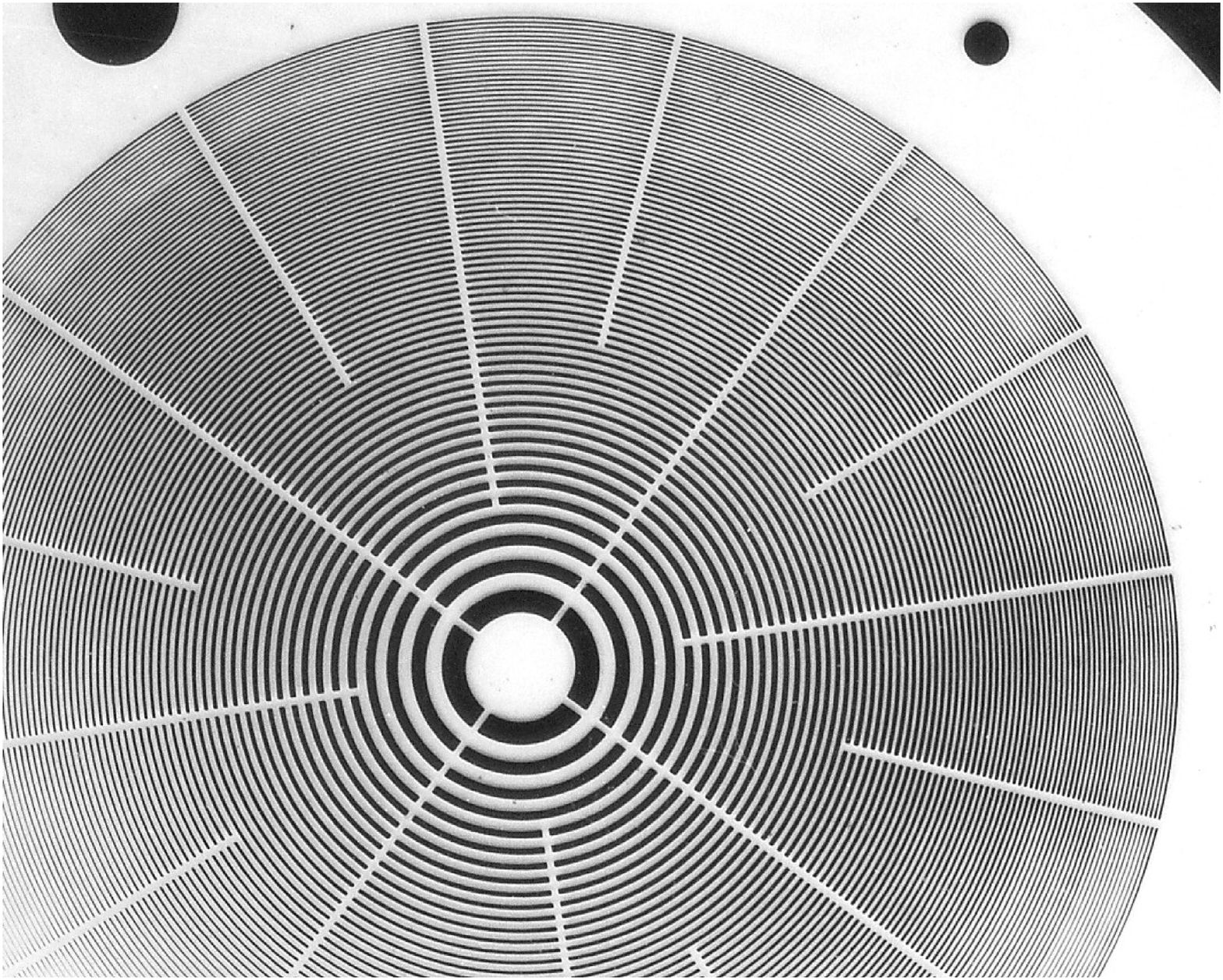}
\includegraphics[height=1.80in]{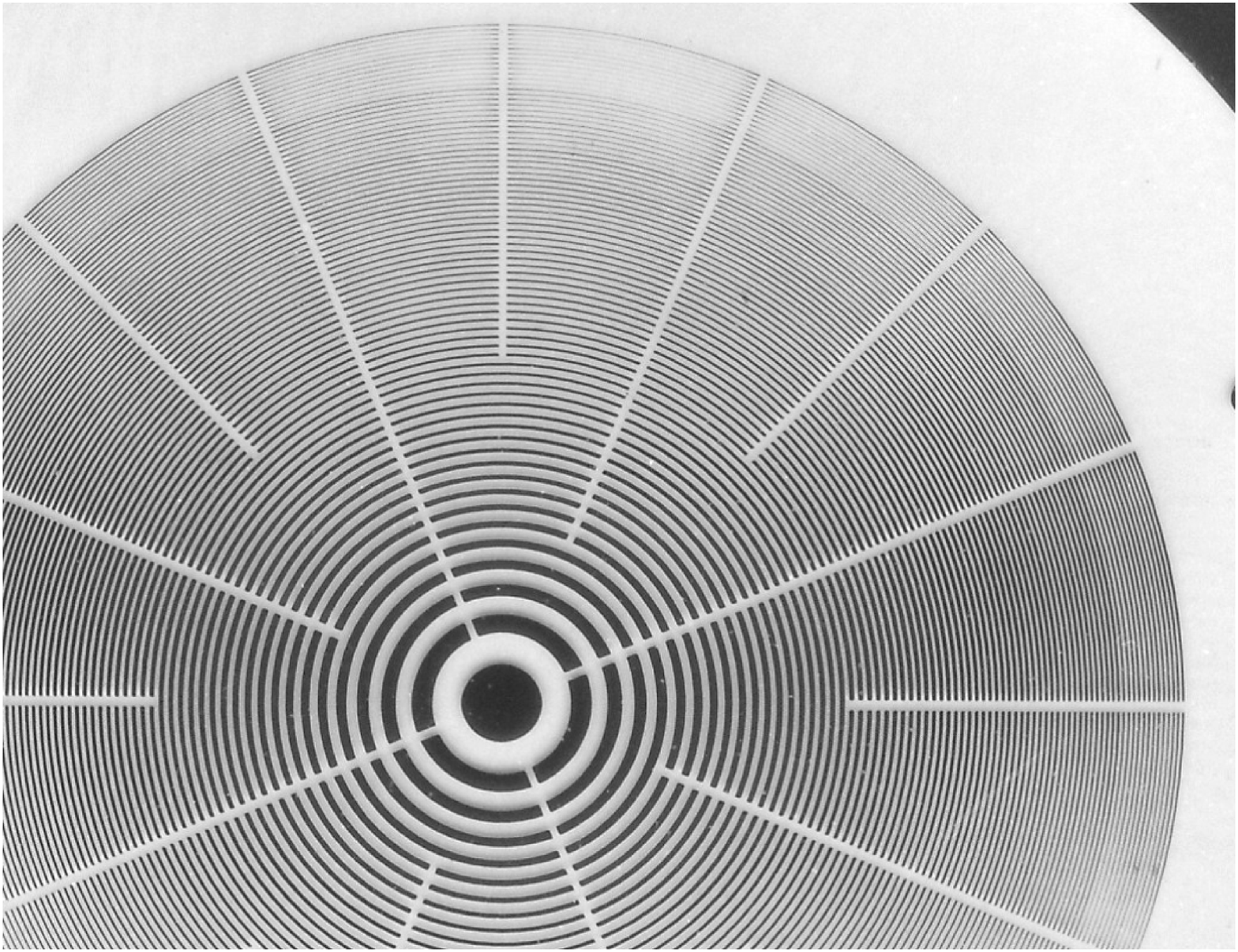}
\caption{Images of two tungsten zone plates on the X-ray film taken at our laboratory. (a) Positive
plate with $r_n = \sqrt{n} r_0$ and (b) Negative plate with $r_n=\sqrt{n-1/2} r_0$. $r_0=0.1$ cm. The finest zones
at the outermost boundary are (a) $41.74$ micron and (b) $41.66$ micron. Angular resolution
(which are independent of the wavelength of X-ray) of a telescope with either of the two of each kind
these plates separated by $20$cm would be $\sim 86$ arc seconds. Radial spokes are to hold the 
circular zones in place.}
\label{}
\end{figure}

\begin{figure}[h]
\includegraphics[height=3.15in]{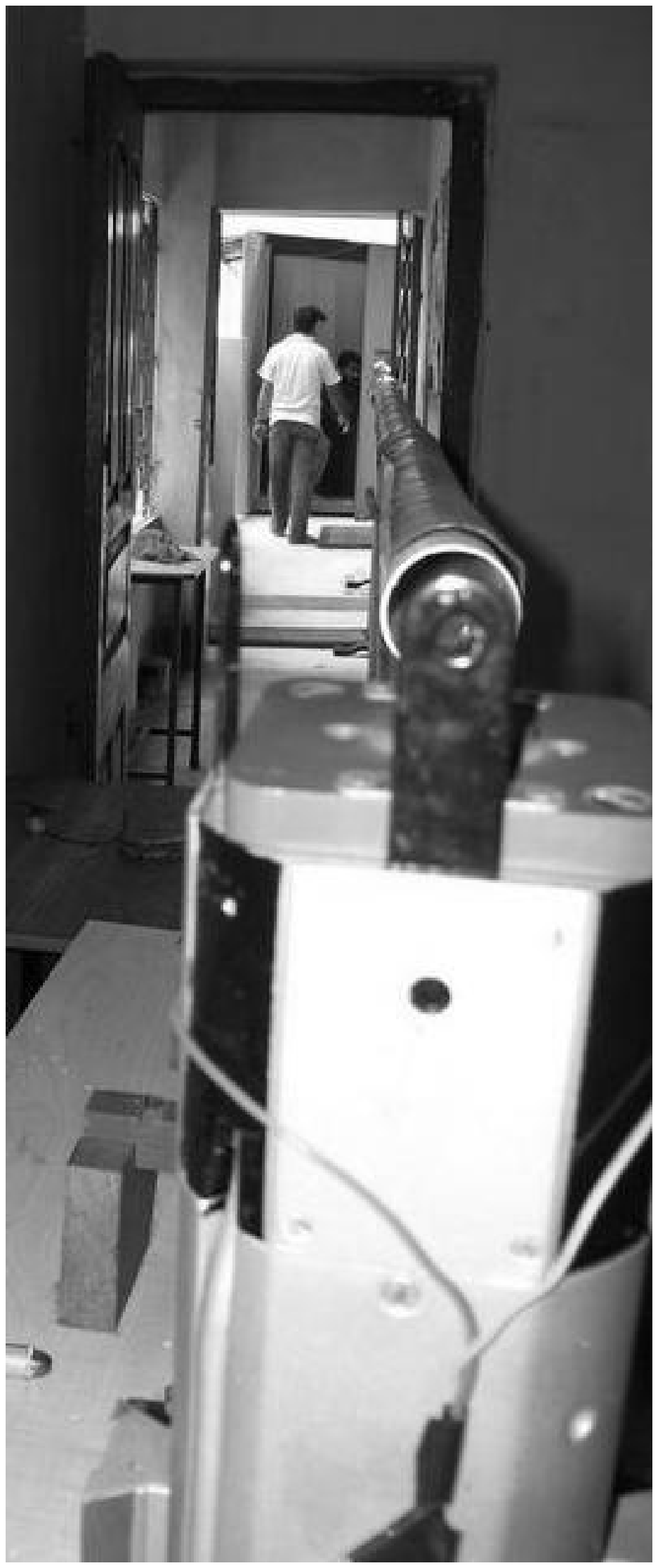} \hspace{0.15cm}
\includegraphics[height=2.65in]{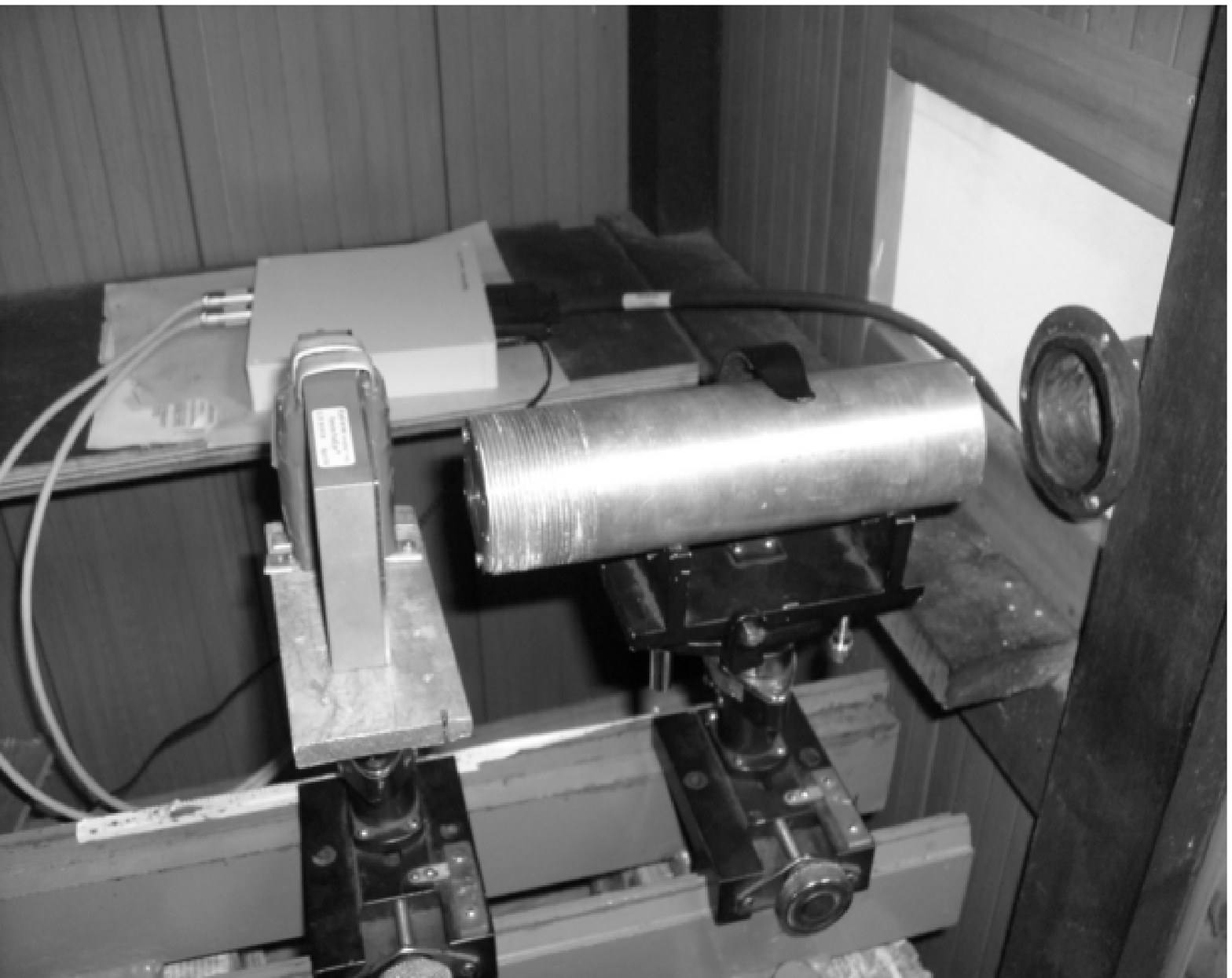}
\caption{(a) The 45 feet beam-line used in our experiments. At the near end is the 
X-ray generator (0-50kV) with molybdenum target and at the far end is the detector assembly.
(b) Set up at the detector end. On the right is the X-ray beam-line. At the center is a $20$ cm 
zone plate holder and immediately on its left is the CMOS detector.}
\label{}
\end{figure}

\begin{figure}[h]
\includegraphics[height=2.15in]{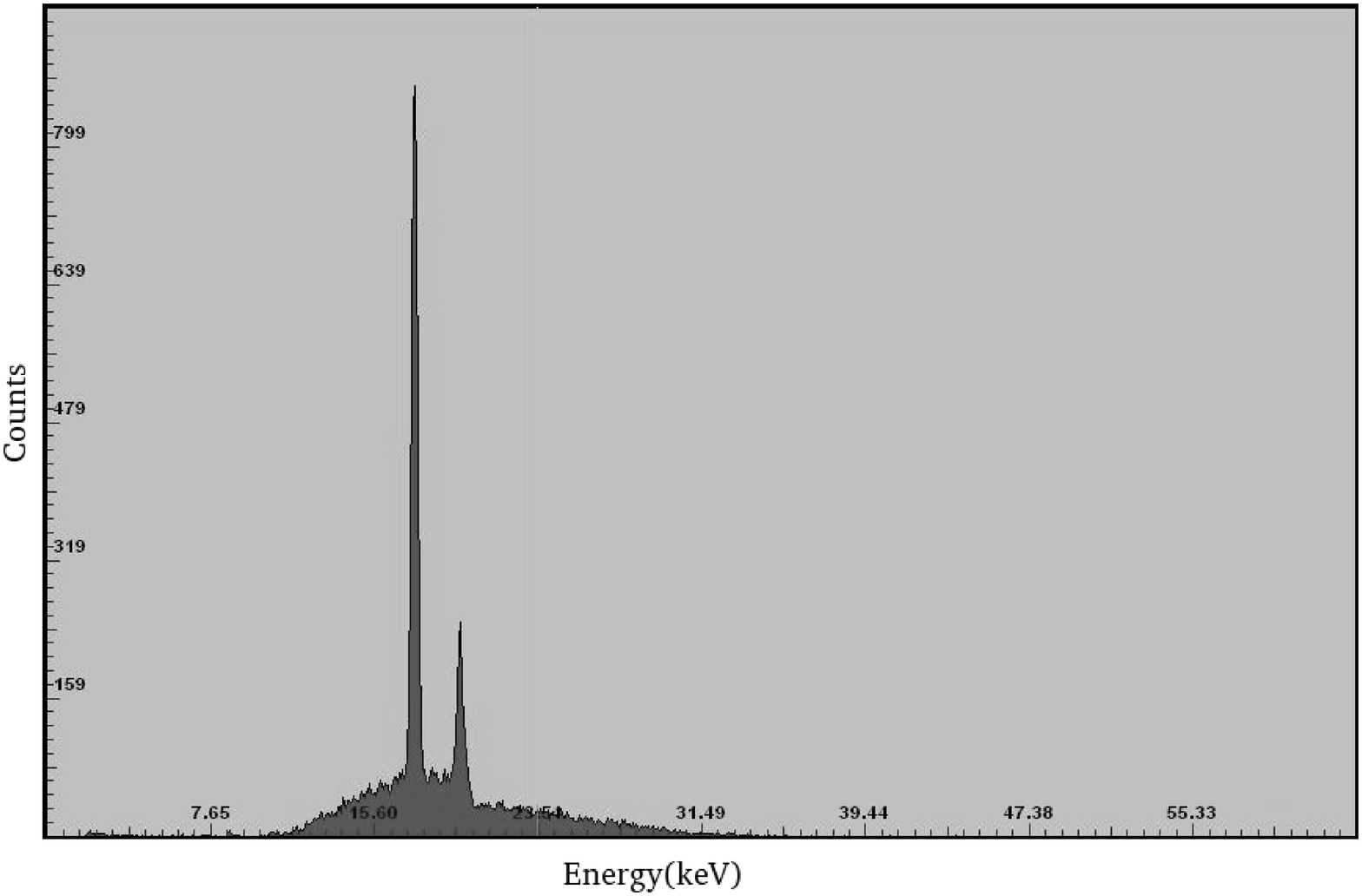}
\caption{Spectrum of X-ray that is falling on the zone plates at the end of the $45$ feet beam-line. 
Energy (in keV) is along the X-axis, and the photon count is along the y-axis. The 
anode voltage is $35$ kV and the anode current is controlled at $10$ mA. The soft X-ray bump in the bremsstrahlung
spectrum is partly absorbed by the intervening air, but the higher energy photons, including 
the $K_\alpha$ and $K_\beta$ lines are roughly as in the original source.}
\label{}
\end{figure}  

In Fig. 4 we show the spectrum of X-ray which reaches at the detector end. The anode voltage was kept at
$35$ kV and the current was controlled at $10$ mA. The typical bremsstrahlung bump is missing due to absorption, 
but the $K_\alpha$ and $K_\beta$ lines are prominent.  

\section{Numerical Simulations}

In a laboratory, very often, it is difficult to change the experimental set-up at will. For instance, if
the detector is placed at a different distance, the whole set is to be aligned once again. 
Similarly, it may be difficult to study with arbitrarily high or low photon flux. Ideal and
theoretically predicted resolution may be difficult to achieve experimentally. 
In a laboratory, generally a single point source is available and thus studying
multiple or extended sourced are difficult.
In order to circumvent these problems, we perform Monte-Carlo simulations of 
what the detector plane would `see' at a given set-up. Thus the effects of 
the flux variation and the divergence of the beam on image quality and resolution
could be studied. In \S 2, we already showed that the fringe patterns contain information
about the distance of the source from the first zone plate. Thus, with suitable
arrangements, to be discussed later in this series, one could reconstruct a three-dimensional
image of the source as well. We process the patterns obtained by our simulations just as we do 
the patterns obtained by our experiments. Thus the images can be reconstructed and studied.

For the simulation, we write an IDL code where the number of photons, the position of
the source(s) and the distance of the source along with other experimental parameters
such as the  zone plate characteristics, distance of their separation, detector pixel size
etc. can be fed as input parameters. The fringe patterns are then inverse Fourier
transformed to obtain the source patterns. Details of the simulations under various conditions
will be presented in Paper II (Palit et al. 2008) of this series.

\section{Results}

For all the experiments, we fix the anode voltage at $35$ kV 
and the anode current is set at $10$ mA. The CMOS detector is placed at a
distance of $5$ cm behind the zone plate. Also, the zone plate
holder is kept at a distance of $5$ cm from the place where the X-ray
beam-line pipe terminates. Adding the length of the X-ray shielded pipe ($1311$ cm)
and distance from source the front end of the shielded pipe ($27$ cm), the
net distance between the source and the CMOS detector to the source becomes $1368$ cm 
(=$44.85$ ft).

\subsection {On-axis quasi-parallel source}

Fig. 5a shows the shadow cast by two aligned zone plates of Type A when the source is almost
on the axis of the plate holder.  The circular fringes are formed as expected. 
Theoretically, the diameters of the $m^{th}$ dark fringes are located at (see, eq. 11b)
$r_{d,m} = [(2m-1)\pi/A]^{1/2}$ ($m$ is an integer). Substitution of $A$ yields,
$$
r_{d,m} = r_1 [ \frac{z (2m-1)}{2D(1-\frac{D}{2z})}]^{1/2} .
\eqno{(12)}
$$
In our case, $r_1=0.122$ cm, $D=20$ cm, $z=1338$ cm. We get the theoretical values of the radii of the first 
and the second dark fringes at $0.722$ cm and $1.251$ cm respectively. The experimental results 
are $0.708\pm 0.02$ cm and $1.23\pm 0.02$ cm respectively which agree extremely well.
Fig. 5b shows the fringe pattern produced by the Monte-Carlo simulation with $2\times 10^5$
photons. The photons on the detector plane are binned in pixels of size $50$ micron $\times 50$ micron,
exactly same as in the CMOS camera. The outcome, particularly the fringe pattern and the radii, is
identical to the experimental result. 

In order to show that the resolution indeed deteriorates with separation between the 
zone plates, we present the similar results when plates are separated by $D=10$ cm, half as much as used before. 
In Fig. 5c, the experimental results are shown, while in Fig. 5d, the simulations results are shown. The 
theoretical location of the first dark fringe from eq. (12) in this case ($D=10$) happens to be
exactly $1$ cm. This also agrees well with the simulation result. The maximum and minimum of the 
fringes are twice as much broad and thus the resolution is reduced by a half. 

It may be noticed that the centers of the bright fringes in experiments are not exactly located on the
axis. This is because of the absence of a very high precision optical bench in our set-up.
From the shifting of the center of the circle and using eq. (10) above,
it is easy to compute the off axisness of the source. In Fig 5a, it is $19$ arc seconds and in Fig. 5c, it
is $34$ arc seconds. In Figs 5b and 5d, the source was chosen on axis in the 
simulations.

\begin{figure}[h]
\includegraphics[height=2.15in,width=2.15in]{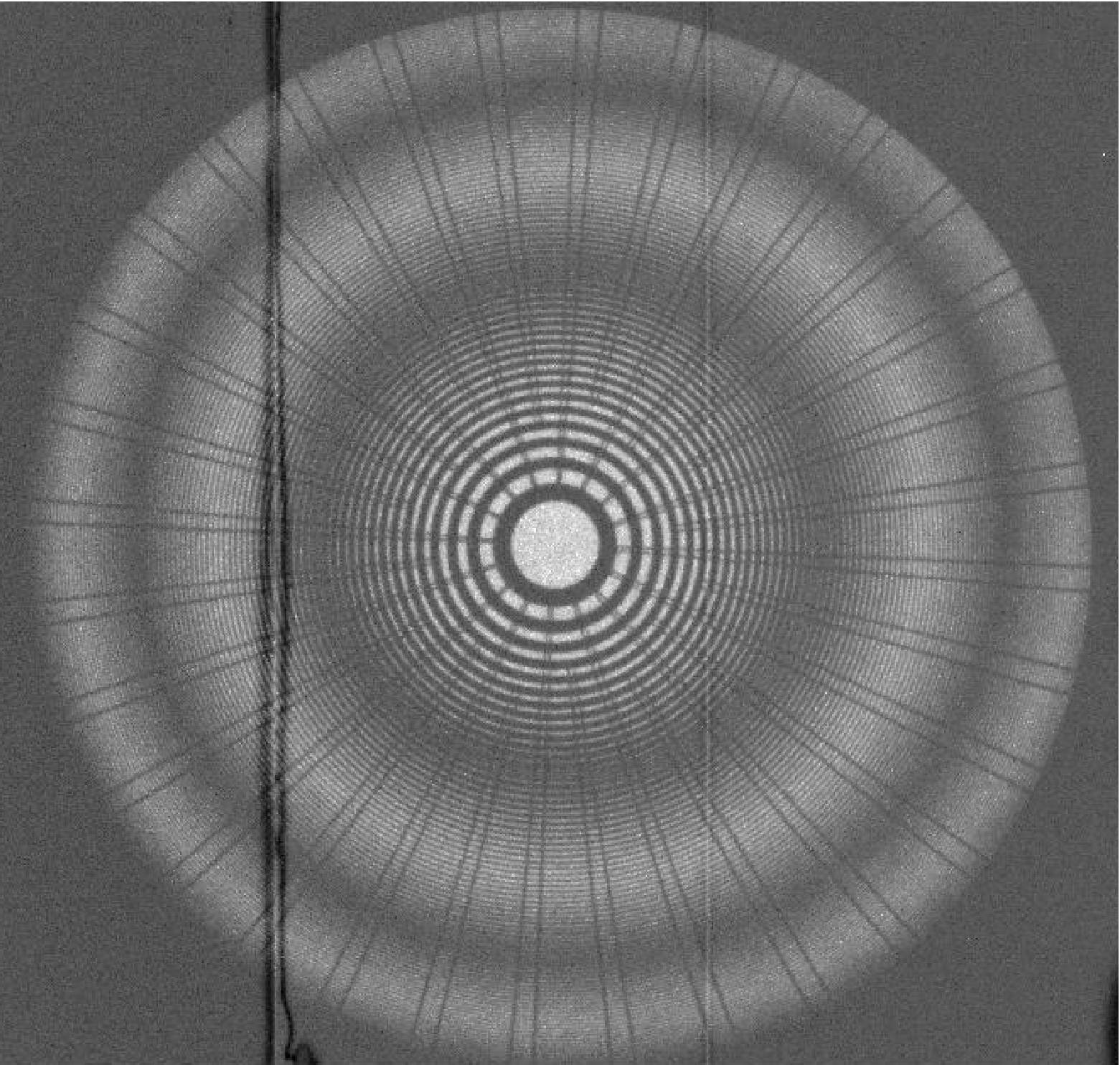}
\includegraphics[height=2.15in,width=3.00in]{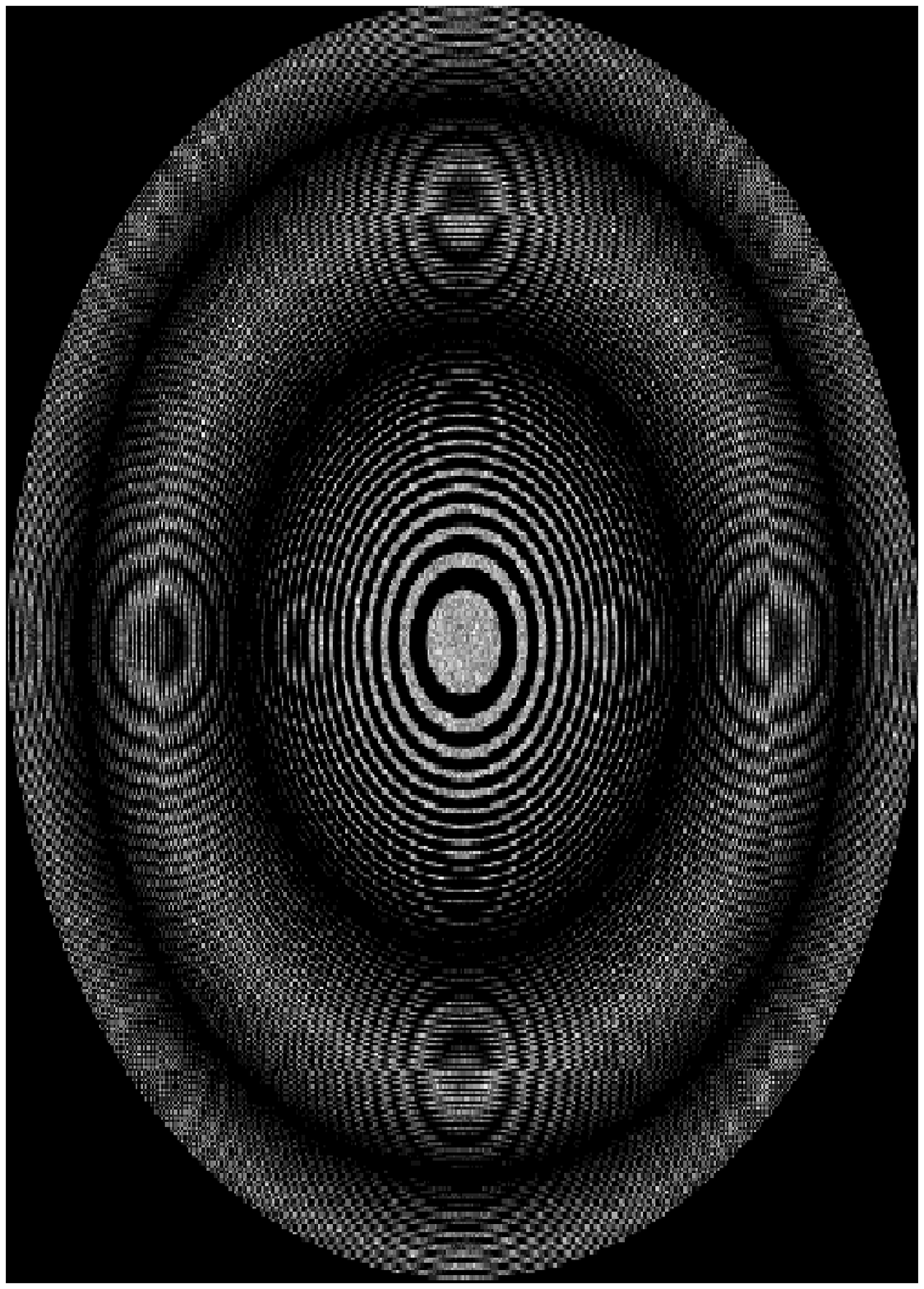}
\includegraphics[height=2.15in,width=2.15in]{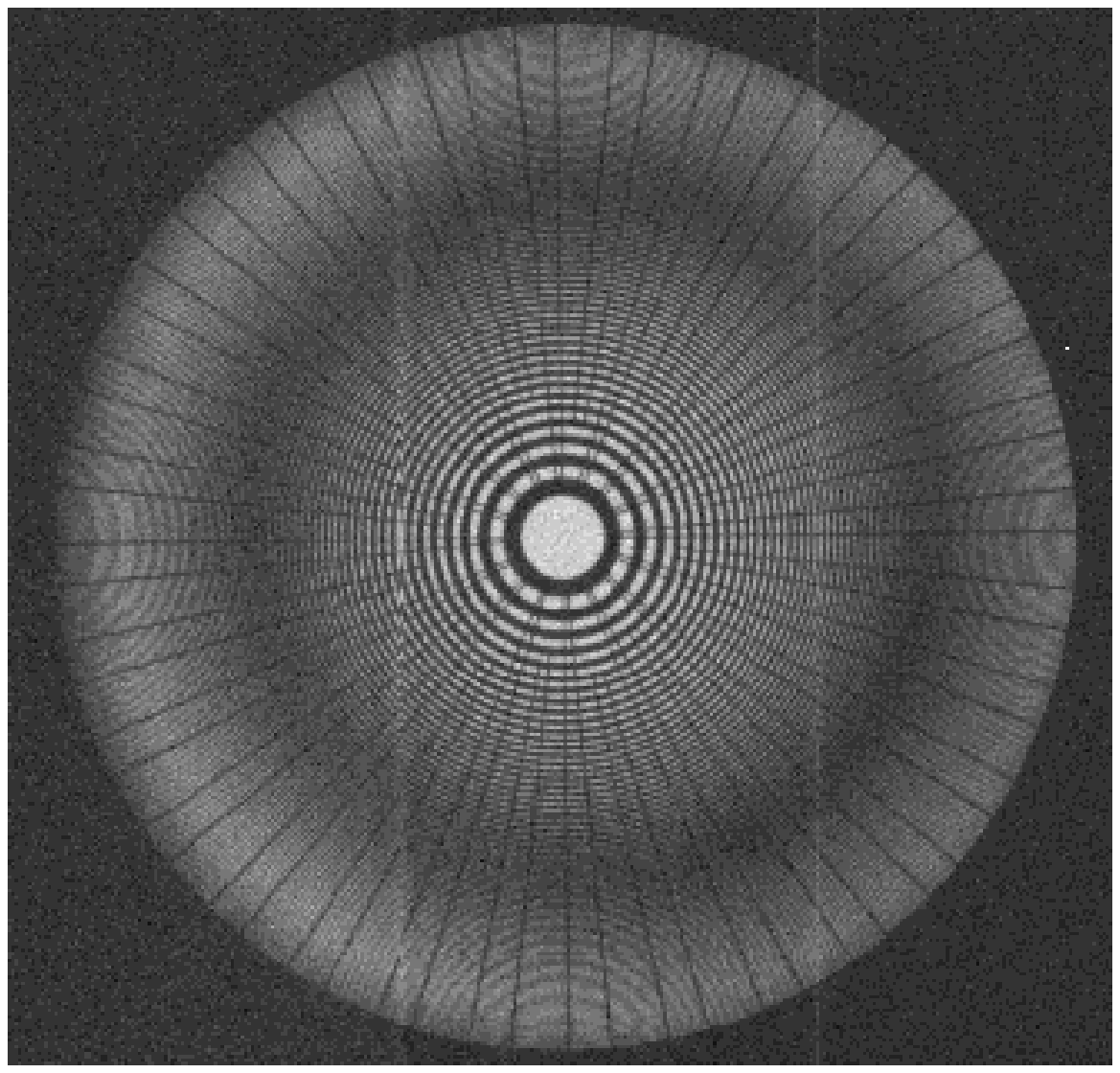}
\includegraphics[height=2.15in]{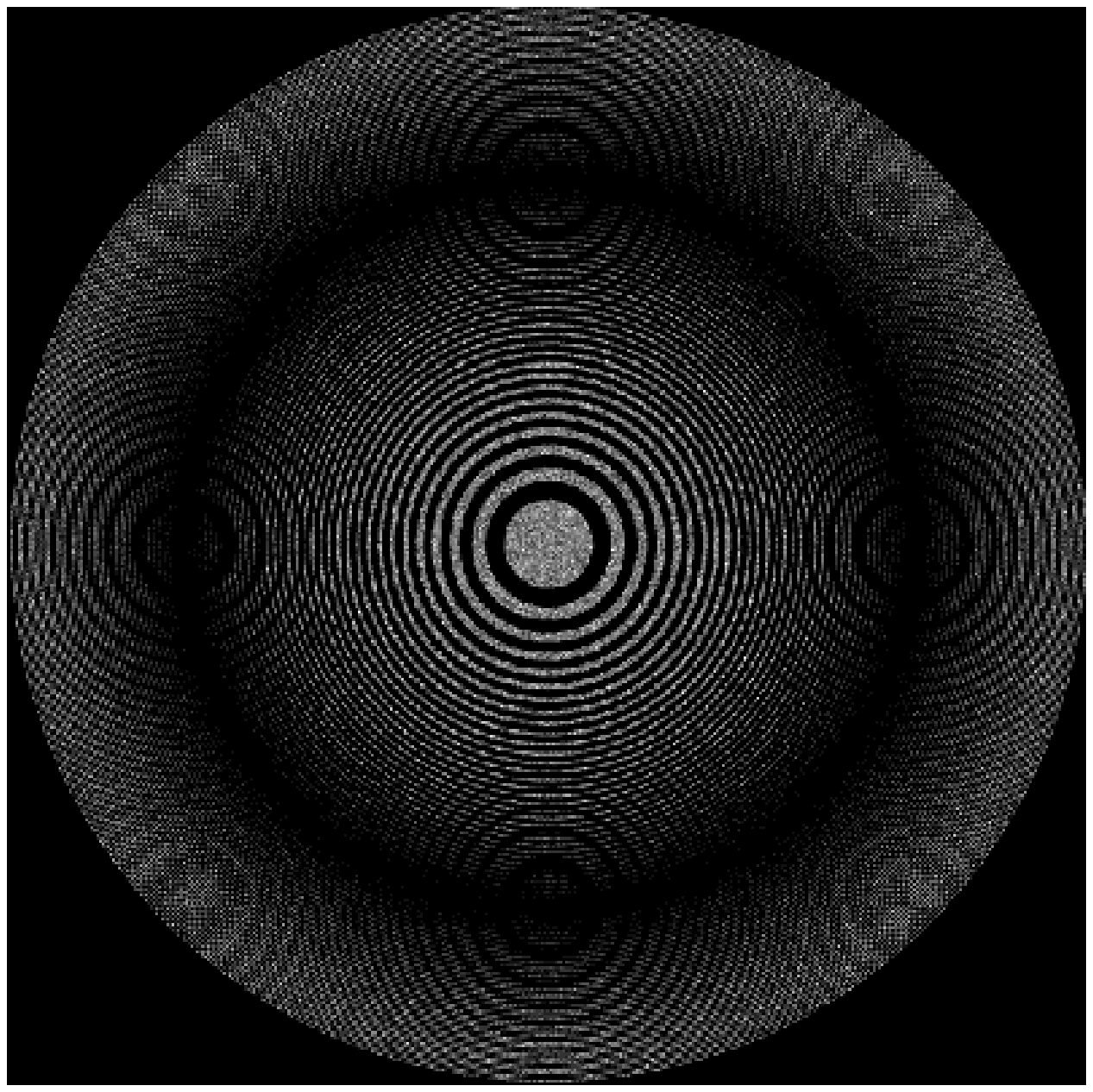}
\caption{Circular dark and bright circles obtained by an on-axis point source located 
45 feet away. Two Type A zone plates were used which are separated by $20$cm in (a-b) and
$10$cm in (c-d). (a, c) Experimental results and (b,d) Monte-Carlo simulation 
results with $2\times 10^5$ photons.}
\label{}
\end{figure}

It is interesting to reconstruct the source by inverse Fourier technique. 
Figs. 6(a-b) depict the on-axis images produced from the experimental and 
Monte-Carlo results presented in  Figs 5a and 5b respectively. The source reconstructed by the
simulation result is somewhat noisy. With a larger number of photons
this problem should go away.

\begin{figure}[h]
\includegraphics[height=1.70in]{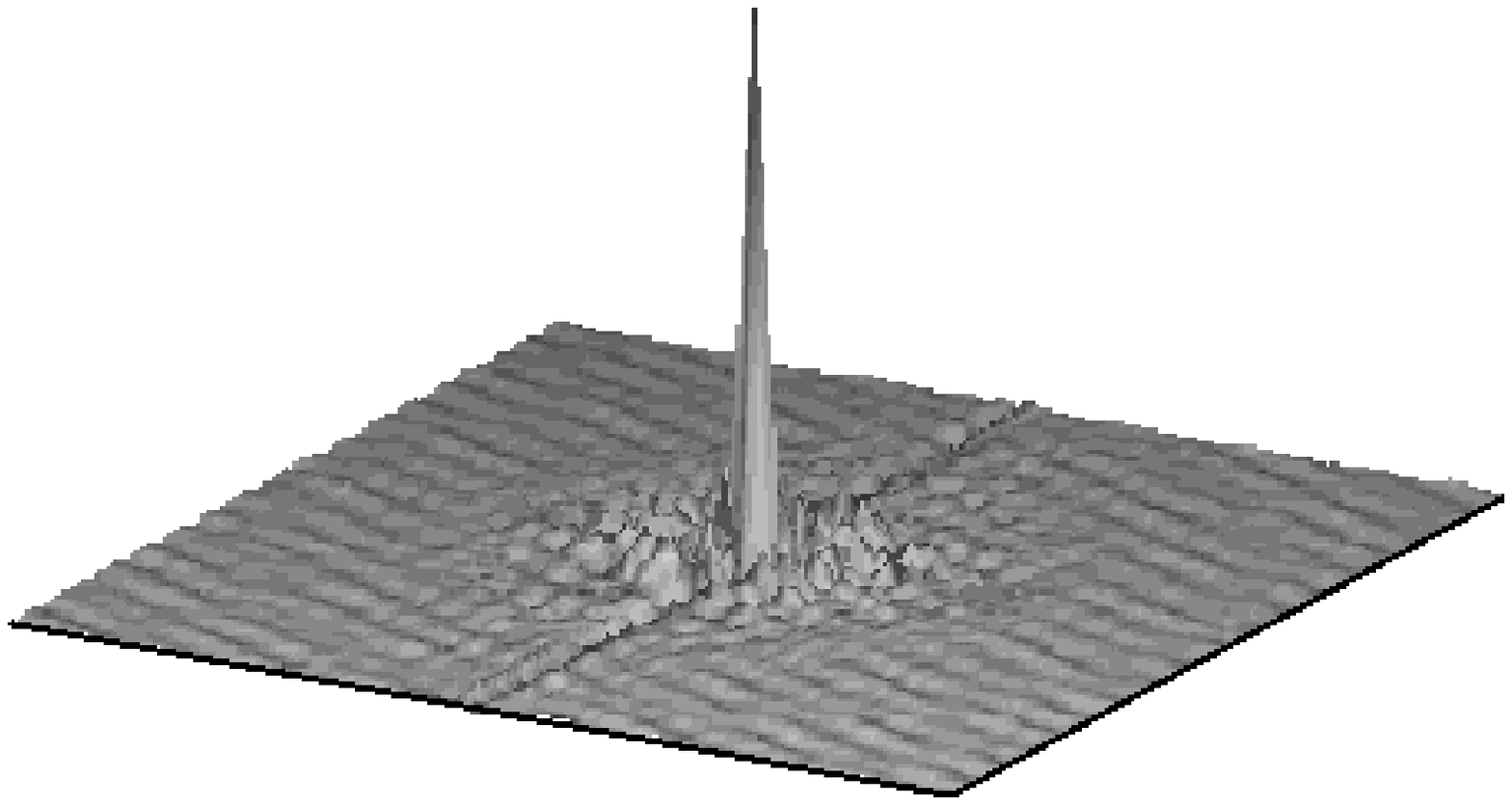}
\includegraphics[height=1.50in]{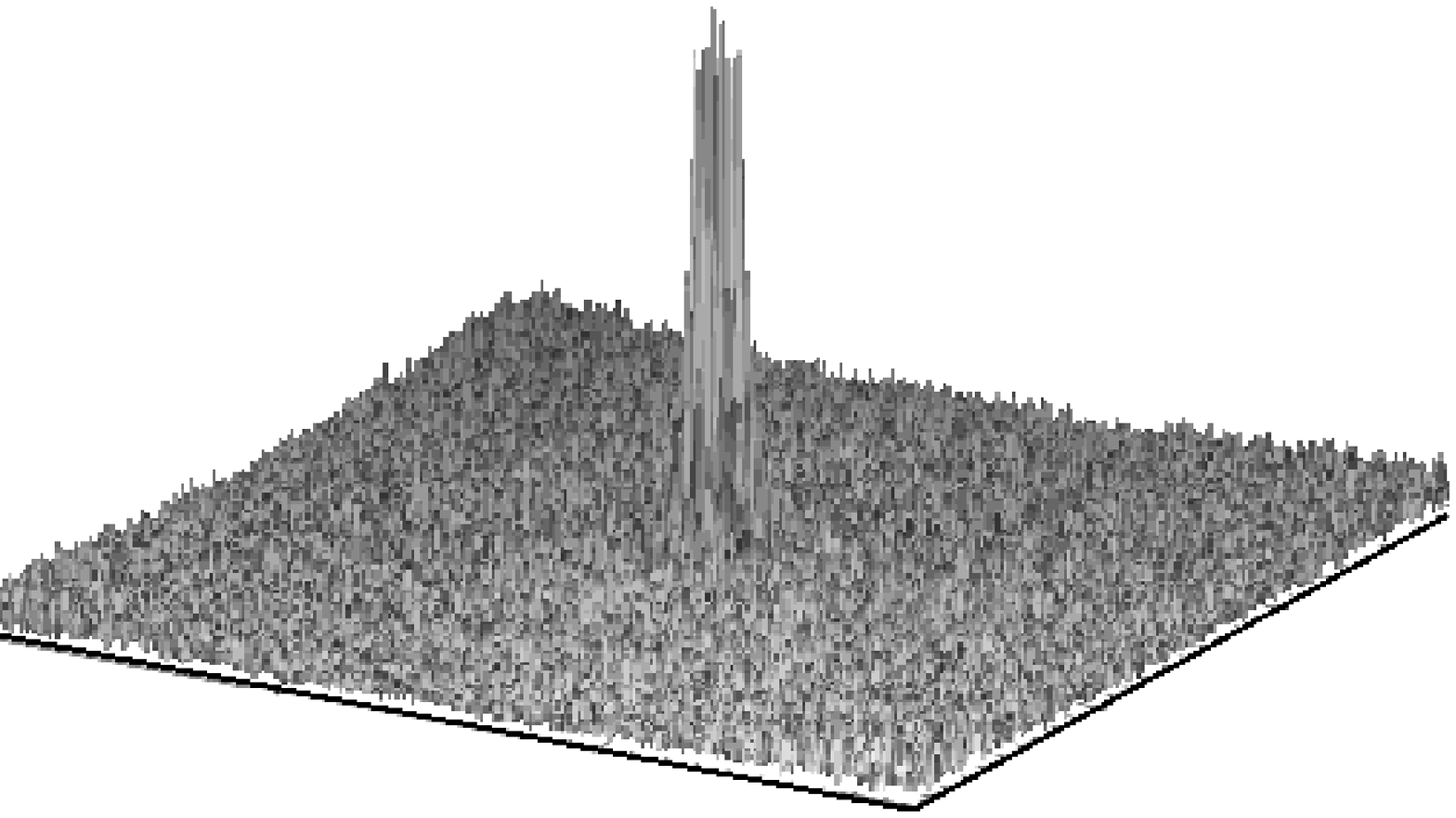}
\caption{
Reconstructed images from the experimental (a) and the simulated (b)
fringe patterns shown in Fig. 5a and Fig. 5b respectively. The 
image by the simulation is somewhat  noisy due to usage of few photons.
}
\label{}
\end{figure}

\subsection {Off-axis single source in a quasi-parallel beam}

We now present the fringe patterns and the reconstructed images when the 
source is off the axis. First we choose the Type A zone plates whose separation is $20$ cm.
We consider the source away from the optical (set up) axis by $0.65$ degree.
The source makes an angle of $15$ degree with the X-axis. 
Figures 7(a-b) show the results of the experiment and the 
of the simulation made with $5 \times 10^6$ photons. The simulated fringe patterns are
in excellent agreement with the experimental results. In Figs. 7c and 7d we show the
reconstructed sources from Figs. 7a and 7b respectively. We note that 
since we use cosine transformers the single source is showing up as a double, symmetrically placed
about the origin. The simulated source is stronger as the photon number employed is high. The source
also has a `ring' around it which corresponds to the point spread function. In reality, each
point source located at a finite distance produces a tiny image of the zone plate. 
As will be discussed in Paper II, this zone plate image starts getting smaller
as the source is placed farther out. Since we 
have used only one transformer instead of four (see \S 2), there is a central pseudo-source
due to the `dc-bias' (the term $1/4$ in transmittance term). This was excised in order to get clear
images. In a complete telescope with four sets of zone plates, such a `dc-bias' would not be 
present due to cancellation of terms. 

In Figs. 8(a-d), we show the similar results of fringe formation and the reconstructed images for $D=10$ cm. Here
the source is shifted away from the optical axis by $0.35$ degree and about $124$ degree away from X-axis.
The experimental results and simulated results show excellent agreement.

\begin{figure}[h]
\includegraphics[height=2.15in]{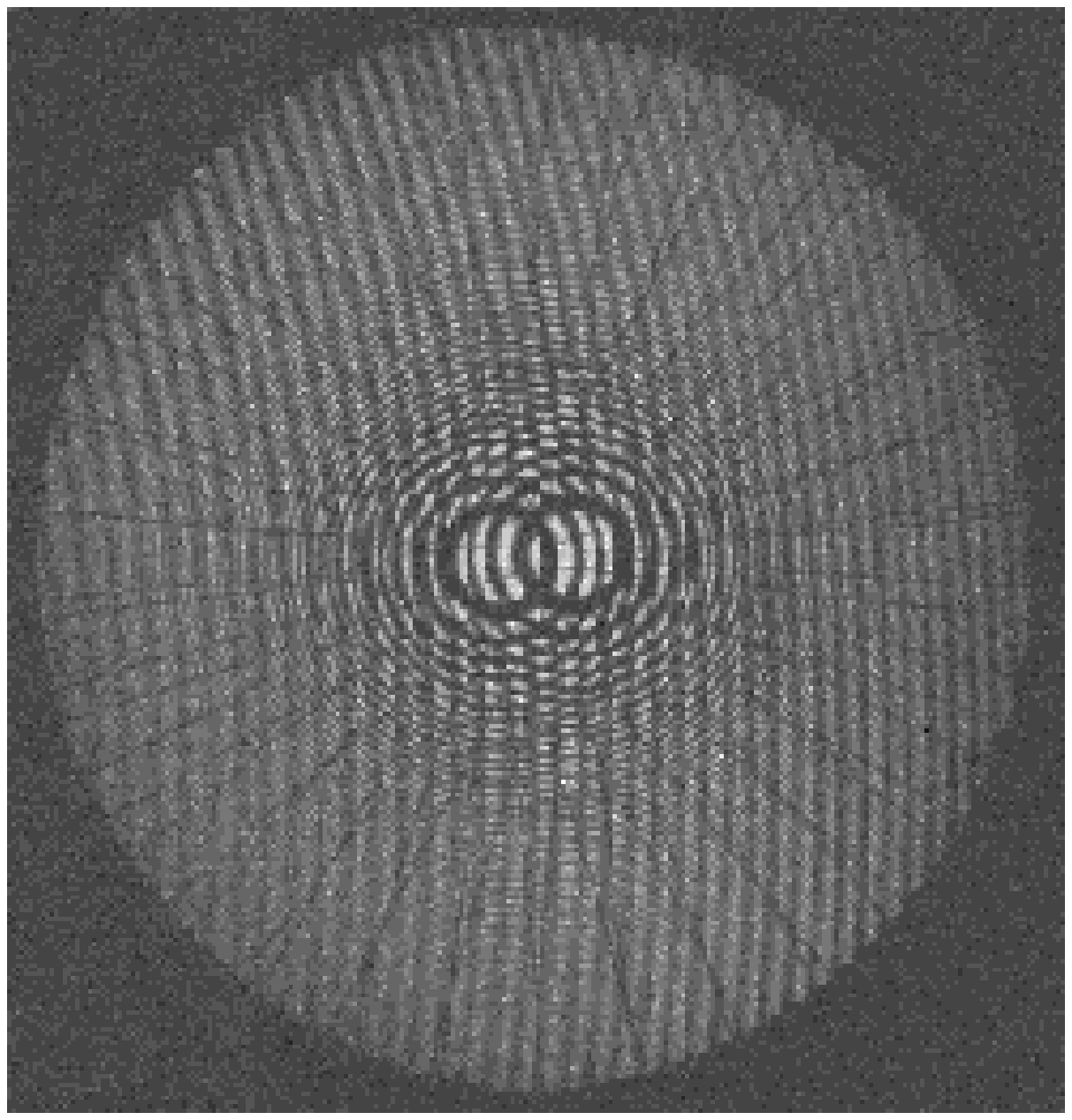} \hspace{0.5cm}
\includegraphics[height=2.15in]{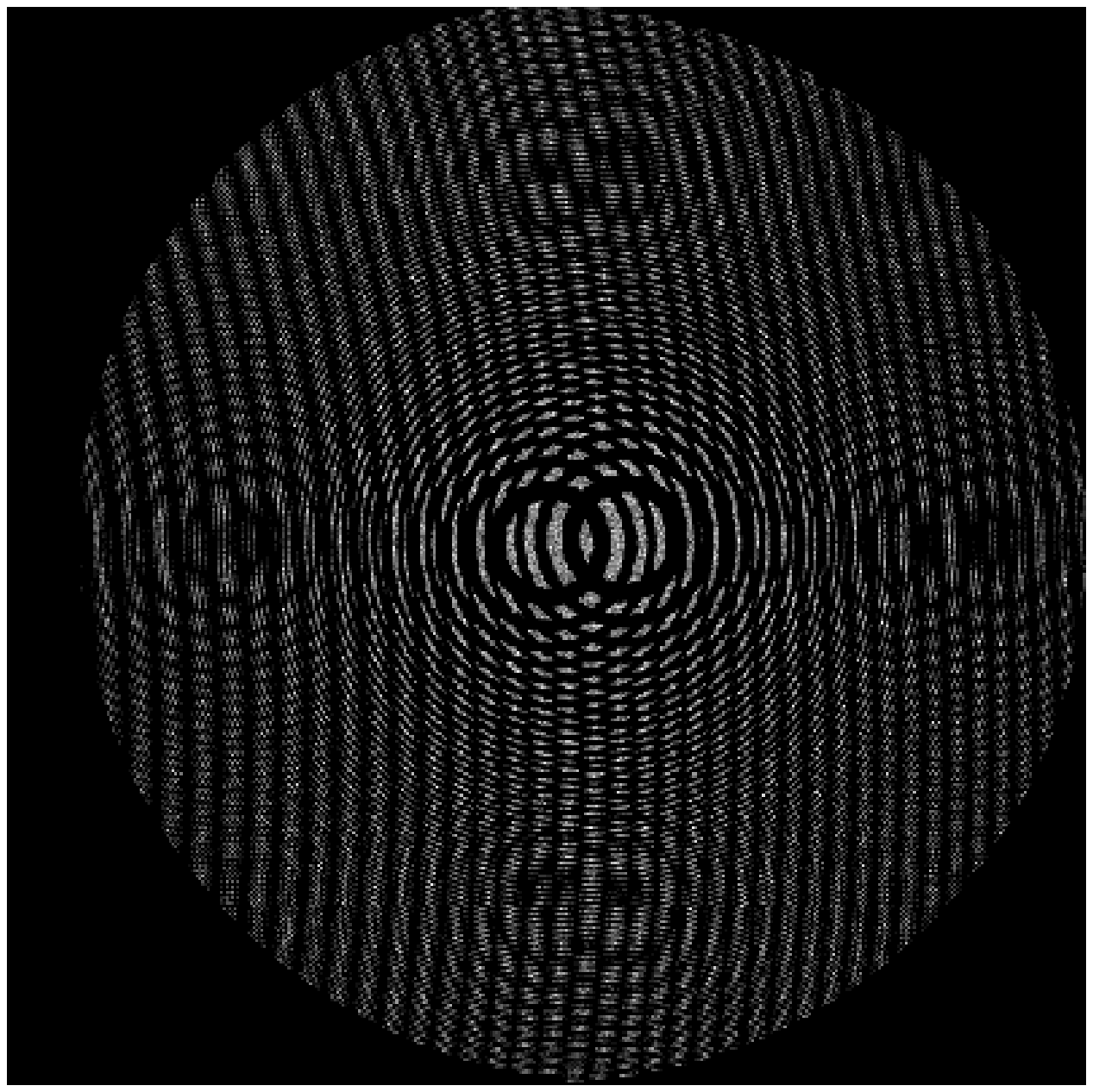}
\includegraphics[height=2.05in]{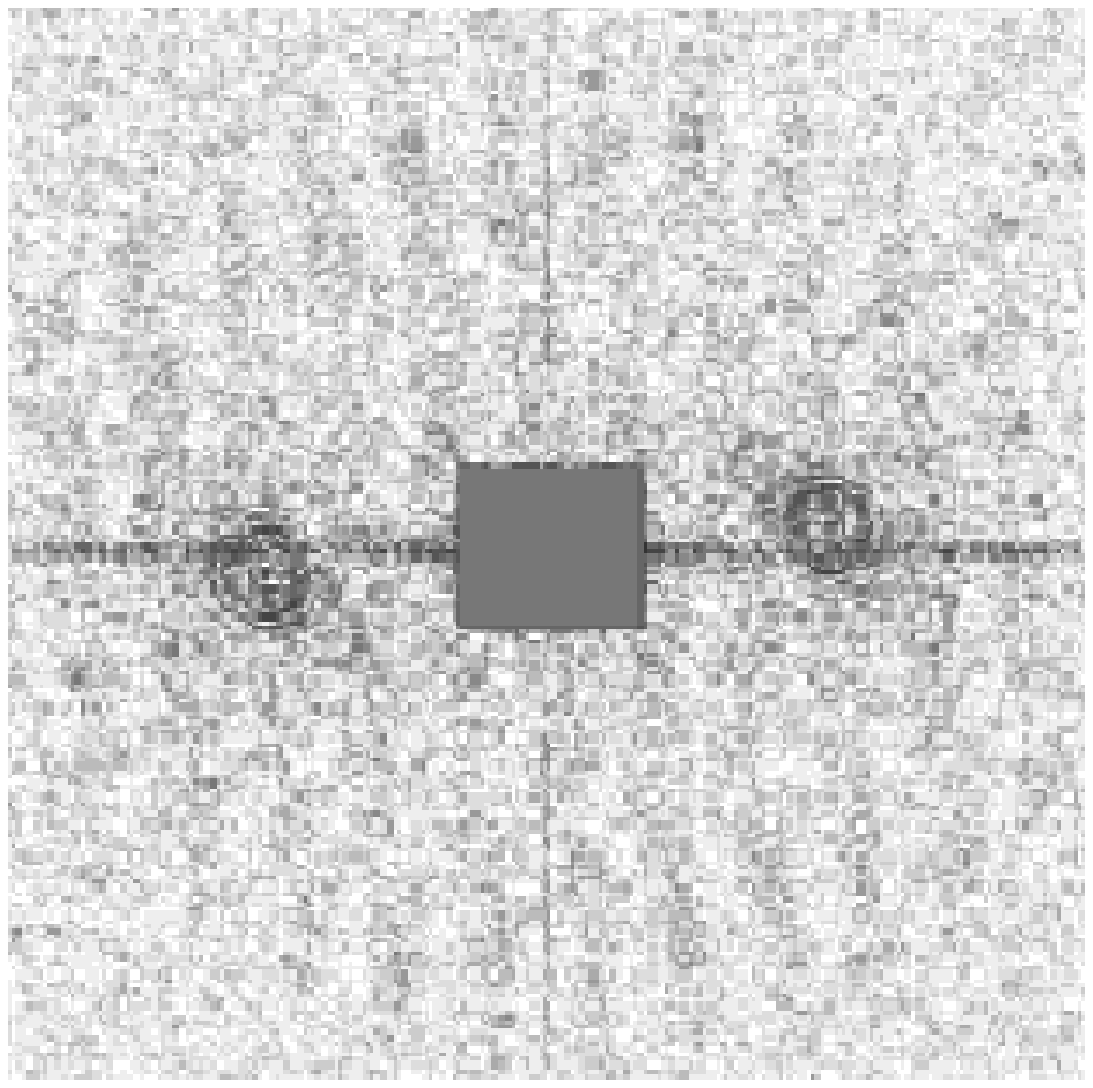} \hspace{0.45cm}
\includegraphics[height=2.05in]{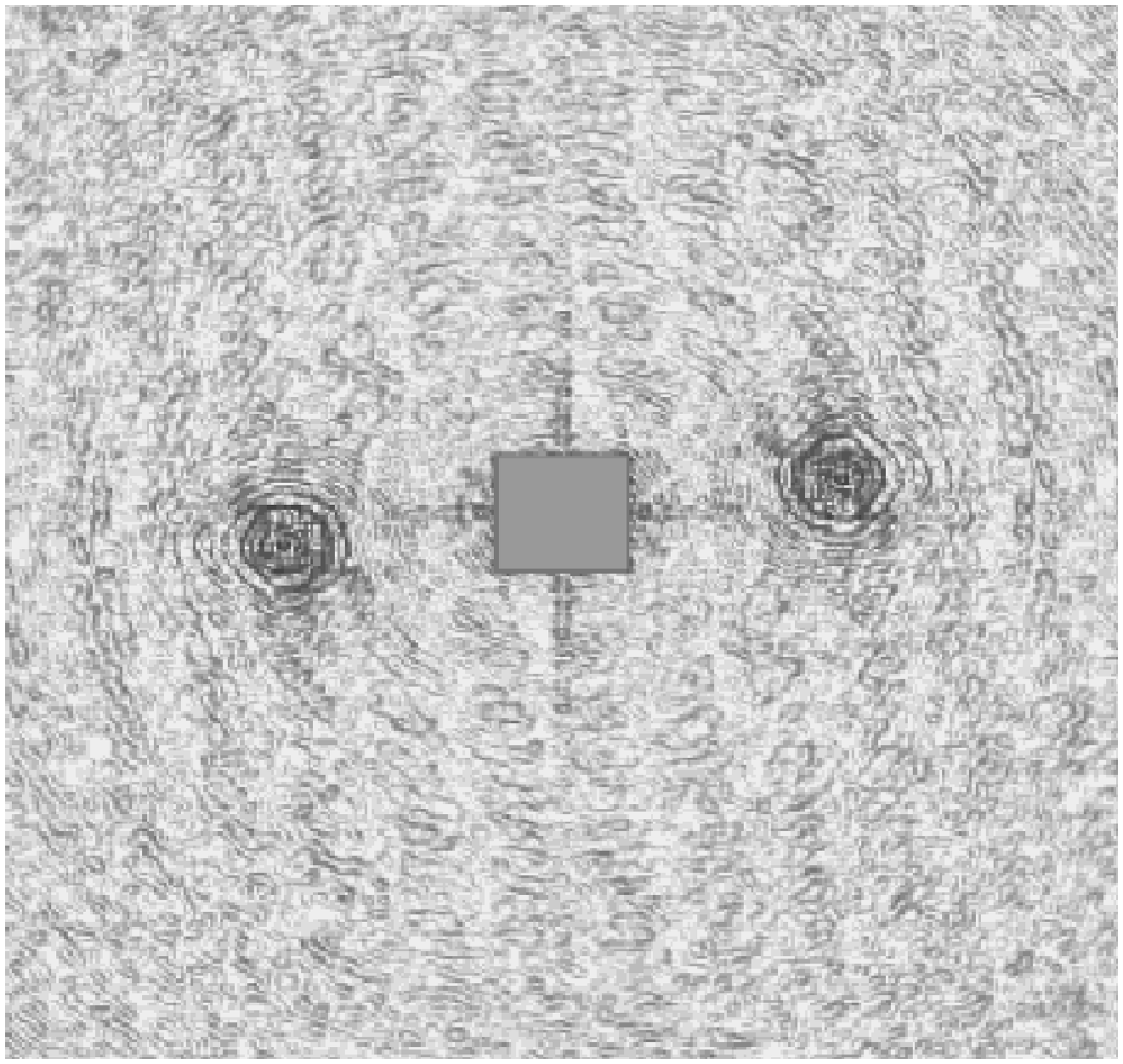}
\caption{Fringe patterns from the (a) experiment and the (b) simulation when the Type A zone plates are 
placed at a distance of $D=20$ cm and the source is $0.65$ degree away from the optical axis. $5 \times 10^6$ 
photons were used in the simulation. The deconvolved images from (a) and (b) are drawn in (c) and (d) 
respectively. They also match, though each dotted image is surrounded by a ring-like point spread function 
obtained after deconvolution. This shrinks as the source is taken farther out.}
\label{}
\end{figure}

\begin{figure}[h]
\includegraphics[height=2.15in]{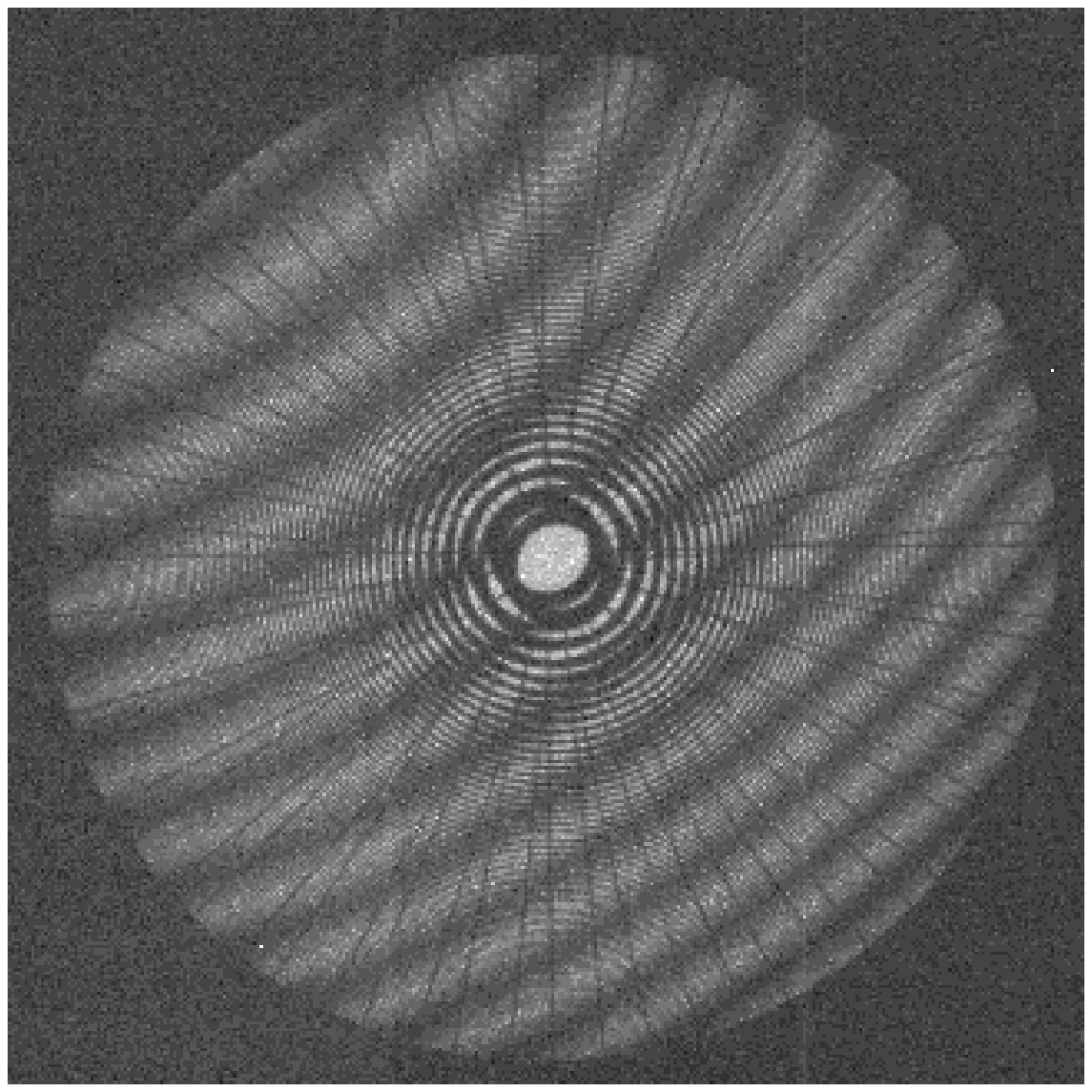} \hspace{0.5cm}
\includegraphics[height=2.20in,width=2.55in]{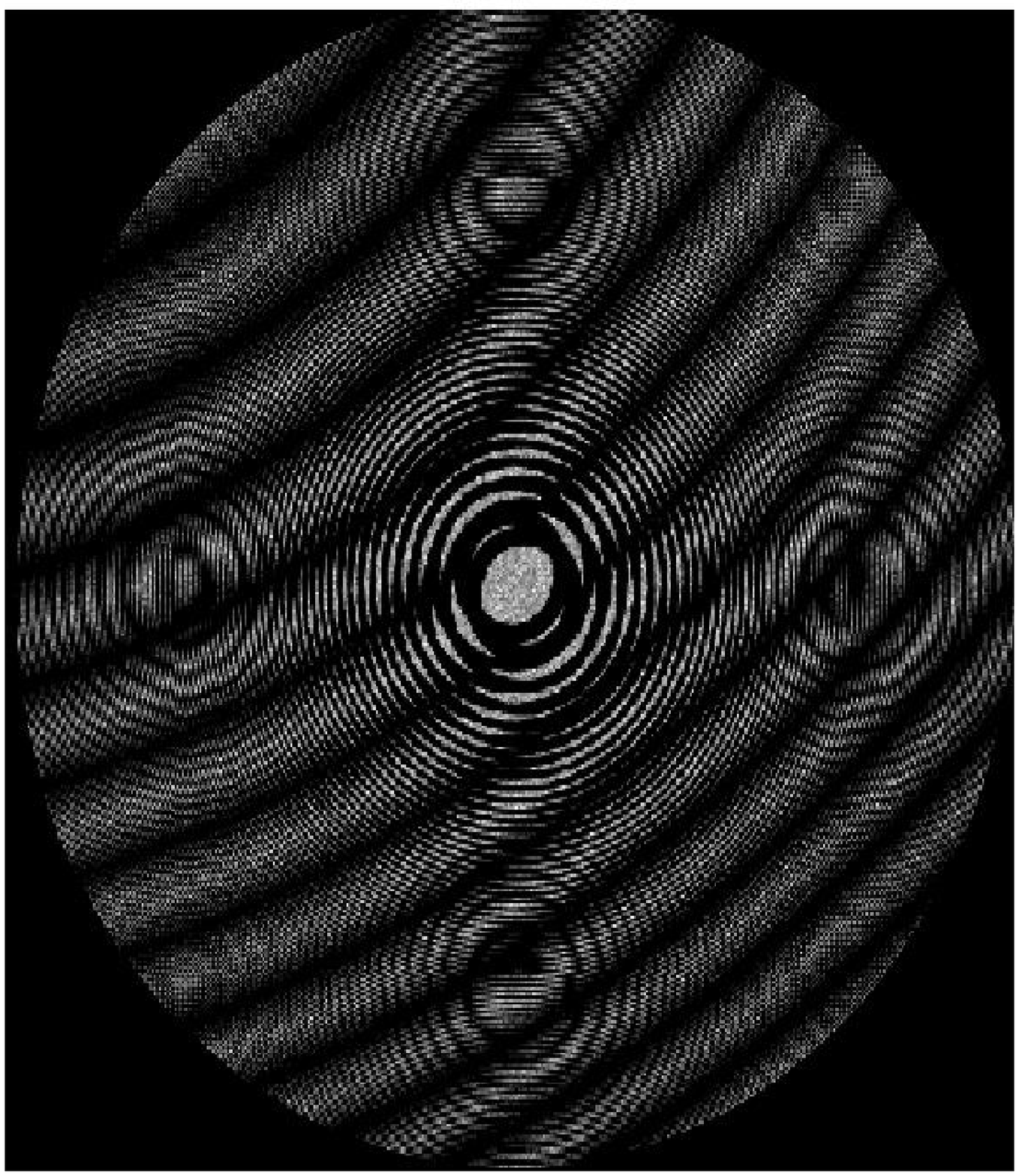} 
\includegraphics[height=2.05in,width=2.15in]{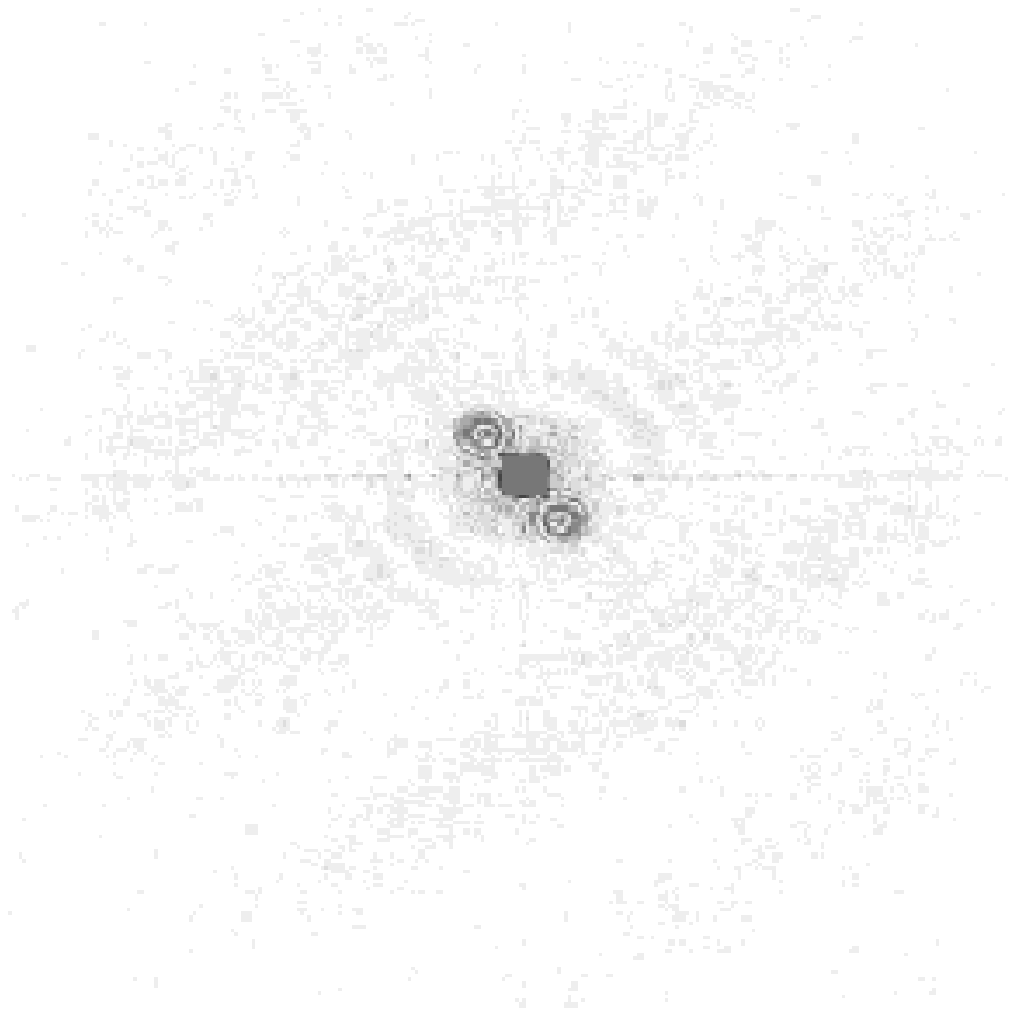} \hspace{0.5cm}
\includegraphics[height=2.05in,width=2.15in]{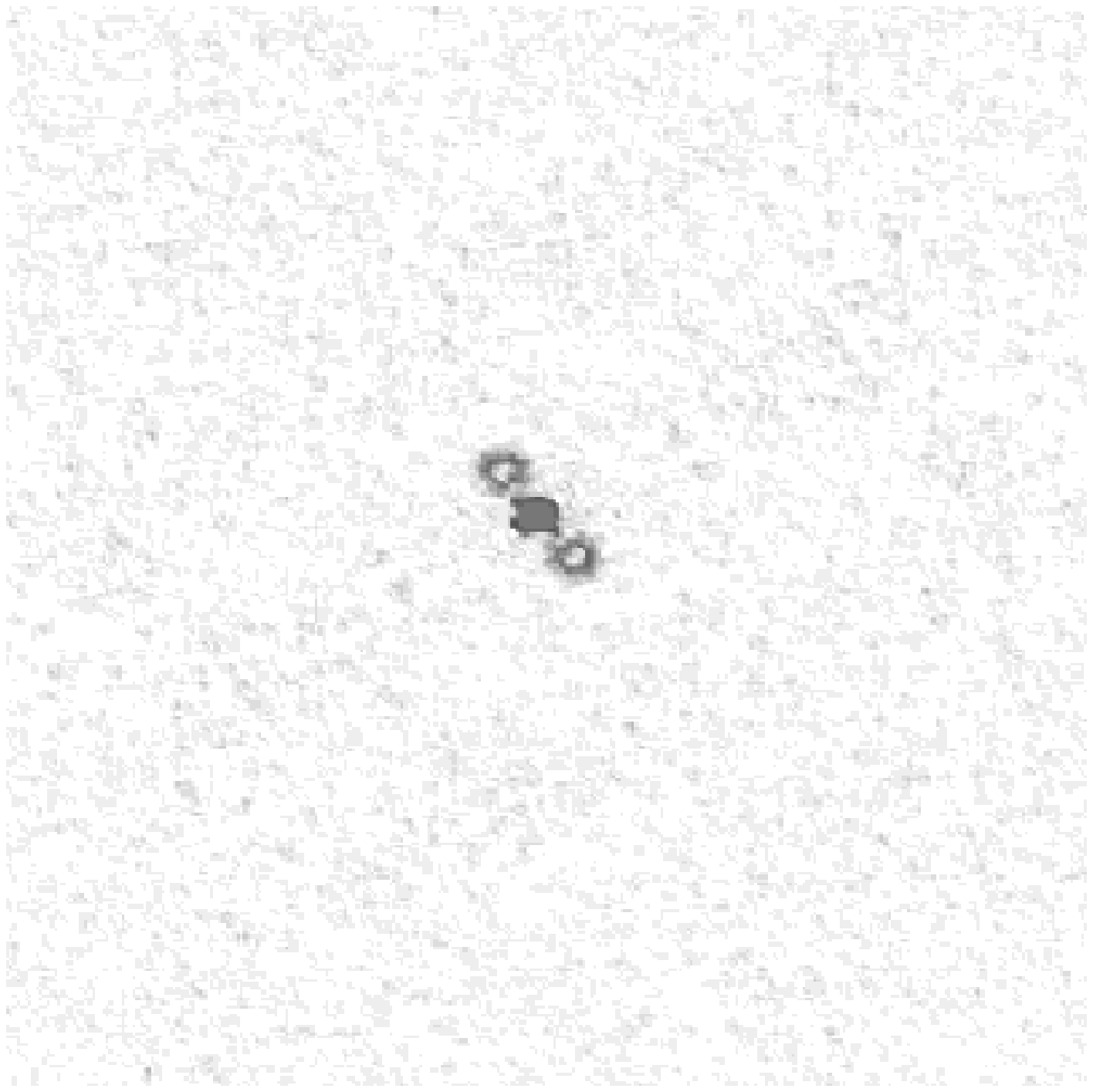} 
\caption{Same as in Fig. 7 except that the distance of $D=10$ cm has been 
used for the zone plate separation. The source is away from the optical axis by an angle of $0.35$ degree 
and and an angle of $56$ degree from the X-axis. $10^6$ photons were used in the simulation. }
\end{figure}

We complete the discussion on an off-axis single source by using two Type C zone plates (one positive and one
negative), so as to verify if the fringes have a phase shift of $\pi$, i.e., the dark fringe 
appears in place of bright fringe, as discussed in \S 2. We place the negative plate towards the 
source and the positive plate towards the detector. The distance between them is $D=10$ cm.
Distance between the detector and the source is $1358$ cm. The source is away from 
the optical axis by $0.7$ degree and the X-axis by $95$ degrees. Figures 9(a-b) show the 
experimental and the simulation results. The central fringe (the fringe passing through the 
intersection of two central zones) is clearly dark. This is to be contrasted with the 
fringes produced by off-axis sources in Figs. 7(a-b) and 8(a-b) which are bright fringes.
Thus the result agrees with our theoretical expectation. The reconstructed images also 
agree very well and we do not give them here.

\begin{figure}[h]
\includegraphics[height=2.15in]{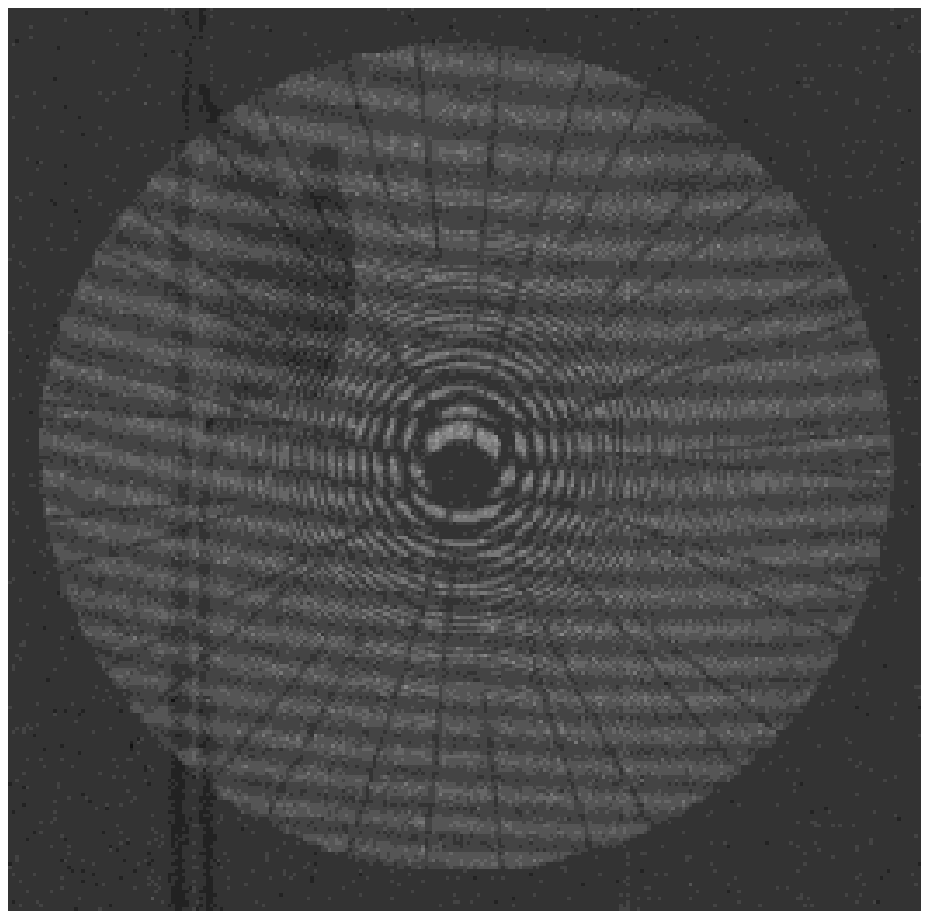} \hspace{0.5cm}
\includegraphics[height=2.15in,width=3.00in]{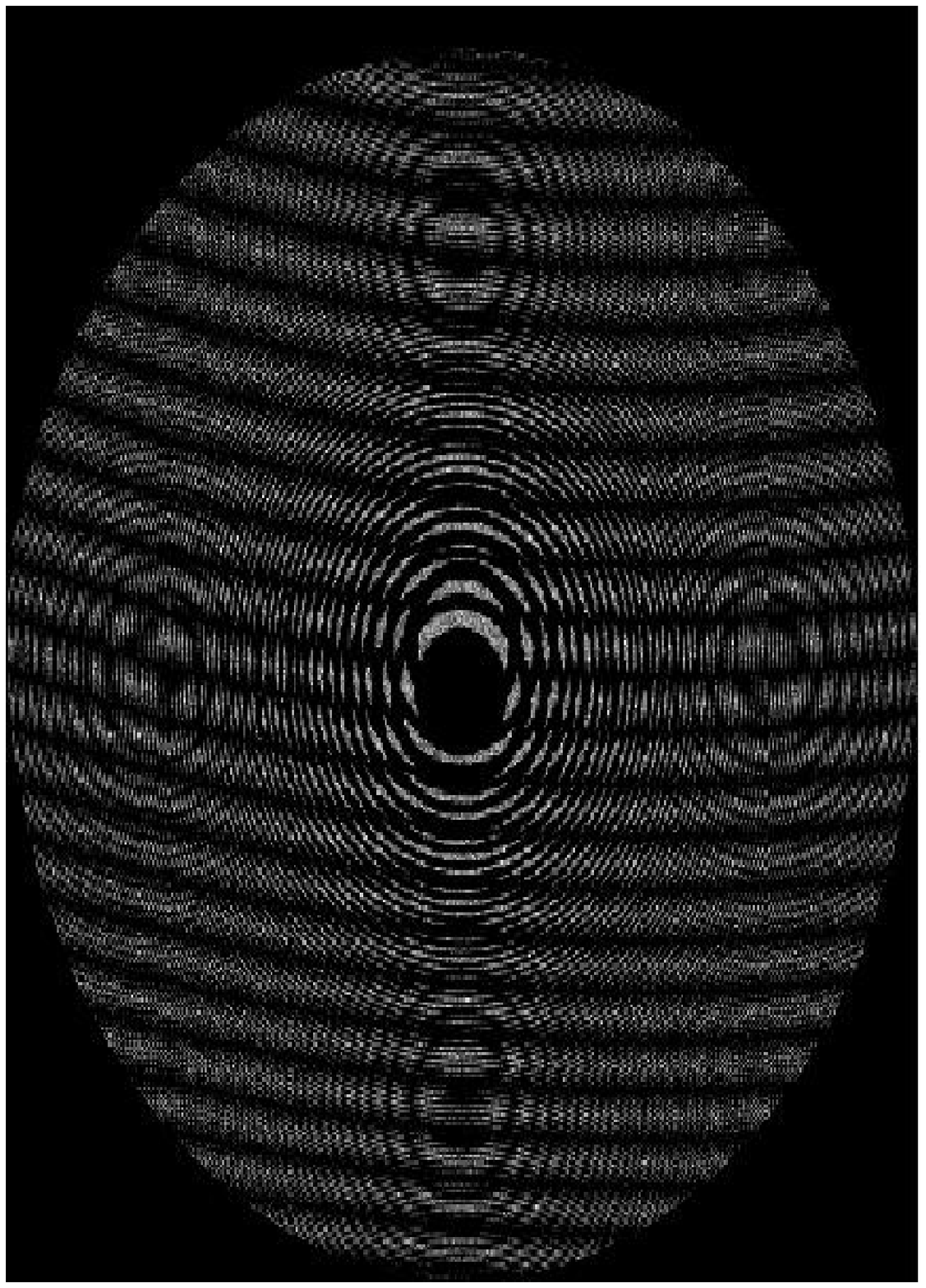}
\caption{ (a) Experimental and (b) simulated fringe patterns of two Type C zone plates
kept at a distance of $10$ cm. One plate is negative which faces the source direction and the 
other plate is positive which faces the detector side. The fringes shifted by a phase $\pi$
with respect to those presented in Figs. 7 and 8.}
\label{exp1-pn}
\end{figure}

\section{Multiple off-axis quasi-parallel sources}

In general, one is confronted with multiple sources within a single frame. In astronomical
situations, such as X-ray sources near galactic centers or flares on the sun could be 
numerous within a single field of view and it is important to resolve them separately. In \S 2, we have 
discussed that with our zone plates, the resolutions are $\sim 80-100 $ arc seconds. 
By choosing a meter-long zone plate holder (which is feasible for any 
medium sized satellite), the resolution can be easily improved to $\sim 8-10$ arc seconds which
is excellent as far as the X-ray astronomy is concerned. 

So far, we discussed the fringe patterns and reconstructed sources when there is a single point
source. Since placing more then one strong source side by side would be difficult, we
can only anticipate how a multiple source image would look like, by superposing a few digital
images from the CMOS detector and reconstruct images out of the composite frame. In Figs.
10(a-b) we present the composite fringe patterns of two sources and the reconstructed
images (including their aliases). Here we used Type A zone plates which are separated
by $D=20$ cm. The sources are away from the optical axis by $0.655$ degree and $0.47$ degree respectively 
and are below the X-axis by $15$ and $14$ degrees respectively. The sources appear as gray circles
corresponding to small zone plate images. In Fig. 10(b), we removed the dc-bias (squared gray area)
as before. The aliases and the central `dc-bias' can be automatically
removed only when four sets of zone plates are used as discussed in \$ 2.

\begin{figure}[h]
\includegraphics[height=2.15in,width=2.15in]{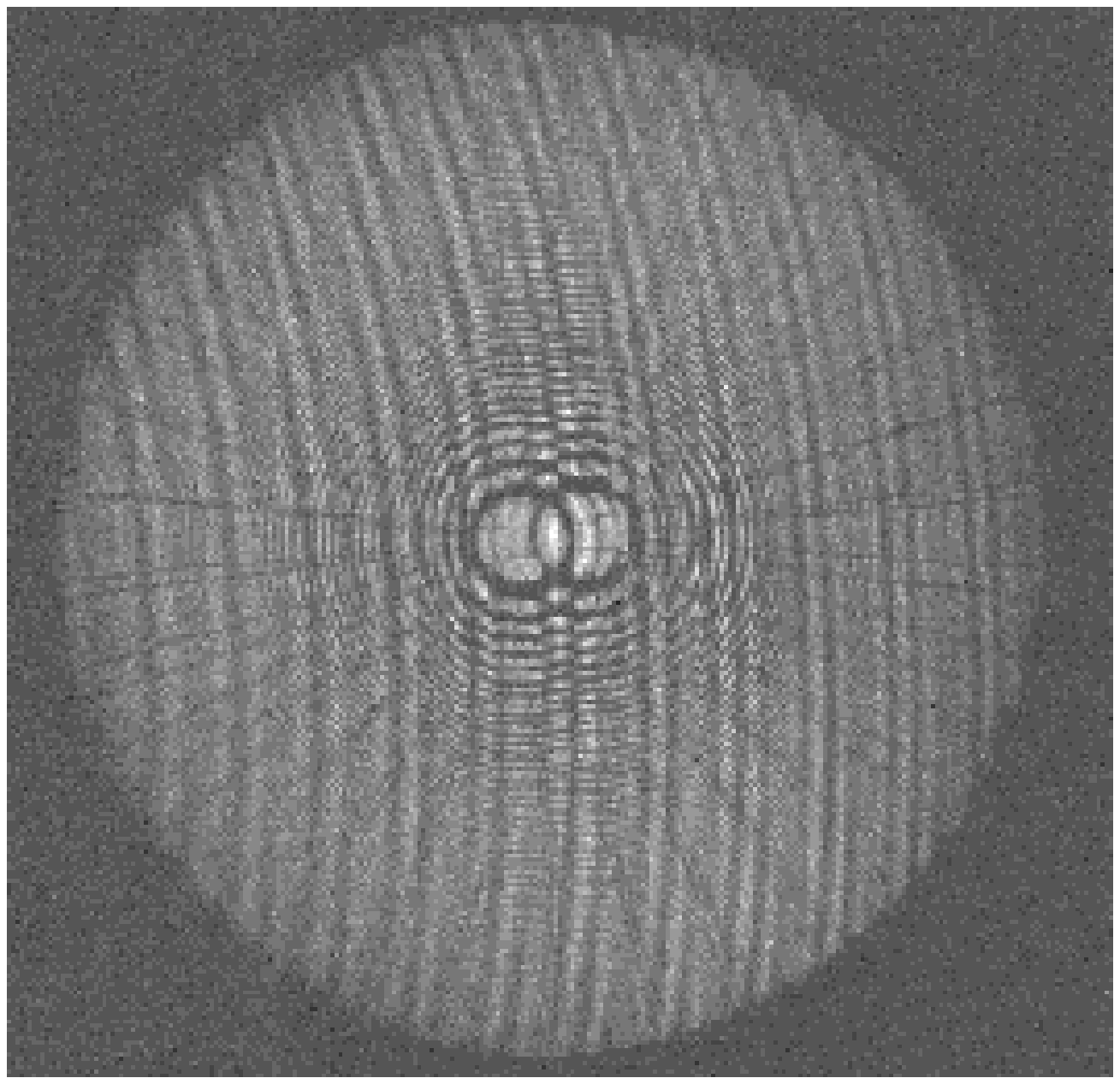} \hspace{0.5cm}
\includegraphics[height=2.15in,width=2.15in]{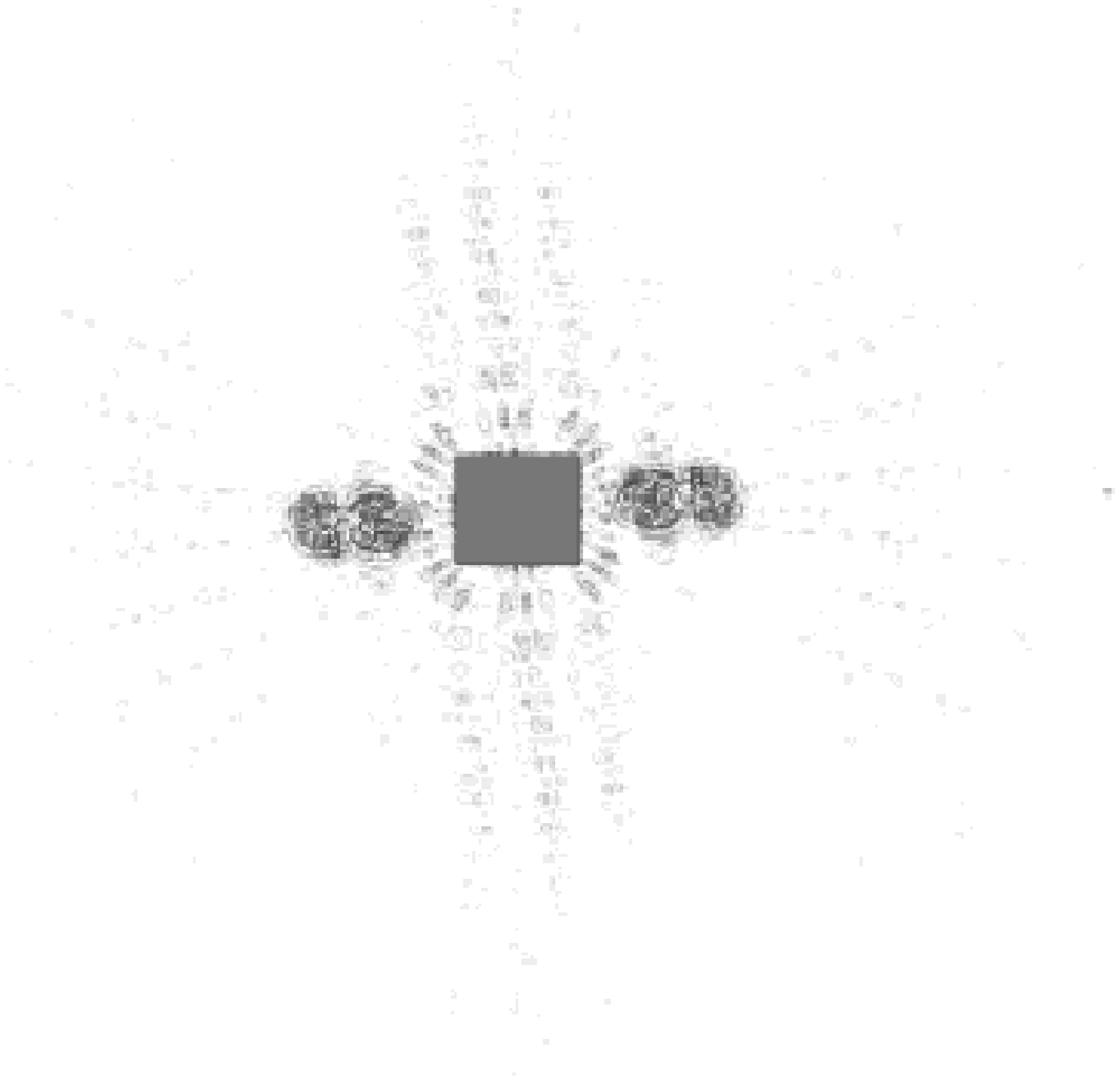}
\caption{ (left) (a) Composite fringe patterns of two sources kept at $0.655$ degree and $0.47$ degree
away from the optical axis of our setup and (right) (b) the reconstructed source distribution. The 
dc-bias [squared gray area] is removed. Since we used only the cosine transformers, 
for every source we have two images symmetrically placed around the origin.}
\label{exp1-pn}
\end{figure}

\section{Discussions and Concluding Remarks}

Zone plates have long been conceived to generate high resolution achromatic X-ray images. Its resolution 
is superior compared to that of the Coded Aperture Masks (CAMs) because of its finer zones. In 
the present paper, we concentrated on the results of an experimental set-up where the X-ray beam
is almost parallel. The collimator is $45$ feet long and a zone plate of $15$ mm radius 
makes an angle of less then $4$ arc minutes at the source. We derived the nature of the 
fringes theoretically and showed that a point source will produce concentric circles
on the detector plane. Through experiments and through Monte-Carlo simulations, we have been able to 
present the fringe patterns and also reconstructed the corresponding images 
both for the on-axis and off-axis sources.
Partly because of the diverging nature of the beam and partly due to the finite size of the detector pixels, 
the ideal theoretical resolution was not achieved. 
This could however be improved during further image processing. 

Resolution of a zone plate telescope could be made very high, since it is inversely proportional
to the distance between the plates. From a practical point of view, in an astronomical satellite,
a distance of a meter is possible which will yield (with smallest zones around $40-50$ micron
as in our experiment) a ideal resolution of $8-10$ arc seconds. 
Another widely accepted type of X-ray imaging instrument having a very high resolution
is the focusing grazing incidence type telescopes which are being contemplated in future space
missions such as NUSTAR and SIMBOL-X. While these
instruments target angular resolutions of $40$ and $20$ arc seconds respectively, 
to achieve these at high energies, formation flight architectures with focal lengths 
of $10$m and $20$m are required. Moreover, grazing incidence telescopes are expensive to 
build and the resolution is energy dependent, even when we assume the pointing 
is very accurate (any grazing type instrument is highly sensitive to alignment). In 
passing, we may remark that at a $20$m separation, our Zone plate telescope will definitely 
achieve an achromatic resolution of less than an arc second.  

One disadvantage of the Zone plate telescopes is that if the size of the finest zone
and the distance of separation of the plates are kept fixed, the resolution will not increase even if the
areas of the plates are made bigger. However, a larger size will gather more photons and will increase the 
sensitivity of the instrument. An achievable zone plate size could be very big (say one meter diameter,
made by low $Z$ metal quoted with high Z metal, such as lead or gold). Even with CZT detectors with 2.5mm
pixel, reasonably high resolution would be achieved.

In order to check the efficiency of a zone plate telescope, we note that with our set up, 
simulations show (Paper-II) that a flux of about $10-12$ photons per cm$^2$ is required to have the source 
resolved about the background. With an increase in area this number does not change! For the crab nebula,
the photon flux in $20-100$ keV range is about $0.12$ counts/sec/cm$^2$. Thus we need to integrate 
about $100$ seconds for imaging purpose. However, for a transient source, our instrument would
be more effective since we could perhaps see change in size of the X-ray emitting region (such as 
glowing of the whole accretion disk when the accretion rate suddenly rises). With a larger
field of view (as opposed to the focusing instruments), zone plate telescopes are ideally suited for
imaging transient sources at a high energy. Of course, to eliminate high energy cosmic rays which 
will increase the background we require the usual protective shielding on our instrument.

In the next paper (Palit et al. 2008), of this series we shall concentrate on the image resolution, and 
how it depends on the source distance and the detector details. We shall also demonstrate the capability 
of a complete zone plate telescope which will constitute four sets of zone plates. In Paper III, we 
shall concentrate on the setups used in the RT-2/CZT-CMOS payload aboard the Russian solar 
satellite CORONAS-PHOTON (Nandi et al. 2008).  

\begin{acknowledgements}
The authors are thankful to Dr. V. Girish and Prof. A. R. Rao for collaborative work in RT2 related
experiments and Dr. U. Desai for many helpful suggestions. 
\end{acknowledgements}


\begin{thebibliography}{}
%
%
\bibitem{}
J.G. Ables, ``Fourier transform photography: a new method for X-ray astronomy",  
Proc. Astron. Soc. Australia, 1, 172 (1968)
\bibitem{}
H.H. Barrett and W. Swindell {\it Radiological Imaging: Theory of Image Formation, Detection
and Processing}, Vols. I and II, Academic Press, New York
\bibitem{}
U.D. Desai, J.P. Norris and R.J. Nemiroff, 1993 in "Astroparticle Physics and Novel Gamma-Ray Telescopes",
SPIE, 1948, 75
\bibitem{}
U. Desai, L. E. Orwig, Mertz, L., Gaither, C.C.III and W. Gibson, ``Shadow Mask Telescope for High Energy X-rays,'' in
High Energy Solar Physics: Anticipating HESSI, Vol. 206, p. 284, Eds. R. Ramaty \& N. Mandzhavidze, 
ASP (San Francisco) (2000) 
\bibitem{}
R.H. Dicke, ``Scatter hole cameras for X-ray and gamma-rays", ApJ, 153, L101 (1968) 
\bibitem{}
U. Desai, L.E. Orwig, L. Piquet and C.C. Gaither, ``X-ray telescope for small satellites,'' Proc. SPIE Vol. 3442, p.94,
Missions to the Sun II, Ed. C.M. Korendyke  (1998)
\bibitem{}
L. Mertz, \& N.O. Young, ``Fresnel Transformation of Images"
in "Proc. Int. Conf. on Opt. Instrum. Techniques", ed. K.J. Habell (London: Chapman and Hall), 305 (1961)
\bibitem{} 
L. Mertz, {\it Transformation in Optics}, Wiley, New York (1965).
\bibitem{} 
L. Mertz, {\it Excursions in Astronomical Optics}, Springer (1996).
\bibitem{}
A. Nandi, D. Debnath, S. Palit, S.K. Chakrabarti, V. Yadav, V. Girish, A. R. Rao, Exp. Astron. (in preparation) [Paper -III]
\bibitem{}
S. Palit, S.K. Chakrabarti, D. Debnath, A. Nandi, V. Yadav, V. Girish, A. R. Rao, Exp. Astron. (in preparation) [Paper -II]
\bibitem{}
G. Pereschi, Mem. D. Soc. It., 79, 26 (2008)
\bibitem{}
B.D. Ramsey, Advances in Space Research, 38, 2985 (2006)
\bibitem{}
A.R. Shulman, {\it Optical Data Processing}, Wiley, New York (1970)
\end{thebibliography}


\end{document}